\documentclass[namedreferences]{solarphysics}

\usepackage[hyperref,optionalrh]{spr-sola-addons} 
\usepackage[optionalrh]{spr-sola-addons} 
\usepackage{graphicx}        
\usepackage{color}           

\newcommand{\etal}{{\it et al.}}

\renewcommand{\vec}[1]{{\mathbfit #1}}

\newcommand{\mnras}{  {\it Mon. Not. Roy. Astron. Soc.}}
\newcommand{\nat}{    {\it Nature}}
\newcommand{\pasp}{   {\it Pub. Astron. Soc. Pac.}}
\newcommand{\pasj}{   {\it Pub. Astron. Soc. Japan}}

\newcommand{\solphys}{{\it Solar Phys.}}

\chardef\us=`\_

\newcommand*{\doi}[1]{\href{http://dx.doi.org/\detokenize{#1}}{{\color{blue} DOI.}}}
\begin{document}

\begin{article}
\begin{opening}

\title{Conditions for Coronal Observation at the Lijiang Observatory in 2011}

\author[addressref={aff1,aff2},email={myzhao@ynao.ac.cn},corref]
{\inits{M.Y.}\fnm{M.Y.}~\lnm{Zhao}\orcid{0000-0002-3408-6549}}

\author[addressref={aff1,aff2,aff3},email={lyu@ynao.ac.cn}]{\inits{Y.}\fnm{Y.}~\lnm{Liu}}

\author[addressref=aff3,email={elmhamdi@ksu.edu.sa}]
{\inits{A.}\fnm{A.}~\lnm{Elmhamdi}}

\author[addressref=aff3,email={askordi@ksu.edu.sa}]
{\inits{A.S.}\fnm{A.S.}~\lnm{Kordi}}

\author[addressref={aff1,aff2},email={Zhangxuefei@ynao.ac.cn}]
{\inits{X.F.}\fnm{X.F.}~\lnm{Zhang}}

\author[addressref={aff1,aff2,aff4},email={stf@ynao.ac.cn}]
{\inits{T.F.}\fnm{T.F.}~\lnm{Song}}

\author[addressref=aff1,email={tzhj@ynao.ac.cn}]
{\inits{Z.J.}\fnm{Z.J.}~\lnm{Tian}}


\address[id=aff1]{Yunnan Observatories, Chinese Academy of Sciences, Kunming 650011, China}
\address[id=aff2]{Key Laboratory of Geospace Environment , Chinese Academy of Sciences, University of Science \& Technology of China, Hefei 230026, China}
\address[id=aff3]{Department of Physics and Astronomy, King Saud University, PO Box 2455, Riyadh 11451, Saudi Arabia}
\address[id=aff4]{Chongqing University College of Optoelectronic Enginerring, Chongqing 400044, China}

\runningauthor{M. Y. Zhao, \etal.}
\runningtitle{Sky Brightness Statistics}

\begin{abstract}
The sky brightness is a critical parameter for estimating the coronal observation conditions for solar observatory. As part of a site-survey project in Western China, we measured the sky brightness continuously at the Lijiang Observatory in Yunnan province in 2011.
A sky brightness monitor (SBM) was adopted to measure the sky brightness in a region extending from 4.5 to 7.0 apparent solar radii based on the experience of the \textit{Daniel K. Inouye Solar Telescope} (DKIST) site survey. Every month, the data were collected manually for at least one week. We collected statistics of the sky brightness at four bandpasses located at 450, 530, 890, and 940 nm. The results indicate that aerosol scattering is of great importance for the diurnal variation of the sky brightness. For most of the year, the sky brightness remains under 20 millionths per airmass before local Noon.  On average, the sky brightness is less than 20 millionths accounts for 40.41\,\% of the total observing time in a clear day. The best observation time is from 9:00 to 13:00 (Beijing time). Lijiang Observatory is therefore suitable for coronagraphs investigating the structures and dynamics of the corona.
\end{abstract}
\end{opening}

\section{Introduction}
     \label{sec:intro} 

The sky brightness is a critical parameter when judging a site's potential for direct coronal observations in optical wavelengths. Due to the large difference in the brightness of the solar disk and the Sun's nearby sky, it is difficult to achieve accurate and direct measurements simultaneously during the day-time. This problem was solved by the invention of the Lyot coronagraph \citep{Lyot1930, Lyot1939}. Lyot coronagraphs are expensive to construct and require extensive support infrastructures to maintain a continuous good performance. A compact sky photometer has been constructed by \citet{Evans1948}. The Evans Sky Photometer (ESP) has been used for a long time at various worldwide solar observatories\citep{Garcia1983,Sakurai2002,Labonte2003}. However it is difficult for the ESP to obtain frequent and continuous measurements throughout the course of the day with CCD systems. In order to overcome these limitations, a sky brightness monitor (SBM) was developed for the Daniel K. Inouye Solar Telescope (DKIST)\footnote{Formerly the Advanced Technology Solar Telescope, ATST.} site survey project \citep{Lin2004,Penn2004b}.

A few years ago, the scientists of the Chinese solar physics community had reached a common understanding that they would find an excellent solar observation site in Western China before developing their next-generation large-aperture solar telescope \citep{Fang2011}. Based on the experience of the DKIST \citep{Keil2000}, the SBM was adopted as a fundamental tool of the site survey project in Western China \citep{Liu2012b}. The SBM is designed following the same idea as was described by \citet{Lin2004} and is used for sky brightness measurements at many different high-altitude sites \citep{Liu2011,Liu2012a,Liu2012b}. At the end of 2013, the ground-based coronagraph NOGIS (\textit{NOrikura Green-line Imaging System}, \citep{Ichimoto1999}) moved to Lijiang Observatory at Yunnan in China. It is necessary to estimate the coronal observation conditions. Fortunately, one of our SBMs works at Lijiang Observatory for long-term observations as a normative reference and accumulated large amounts of data in 2011. During the progress of our site survey project, we have developed a fully automated processing approach to deal with the SBM data \citep{Zhao2014}. In the next section, we describe the improvement of our automatic data processing algorithm and the calibration of the instrumental scattered light. In Section 3, we calculate the statistics of the sky brightness at Lijiang Observatory in 2011. Our conclusions are presented and highlighted in Section 4. 

\section{Data Reduction} \label{sec:DataRe}
\begin{figure}    
\centerline{
\includegraphics[width=0.495\textwidth,clip=]{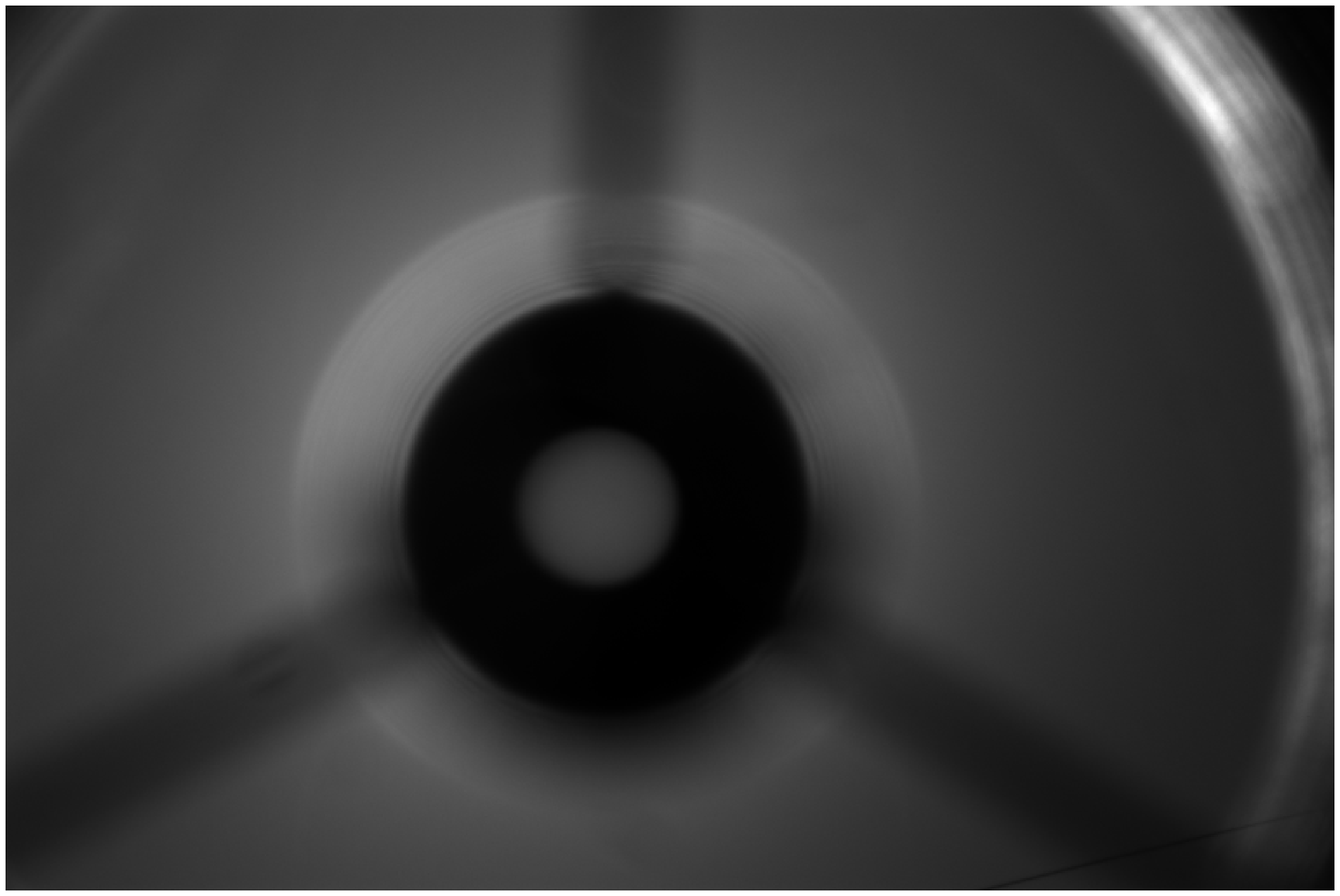}
\includegraphics[width=0.495\textwidth,clip=]{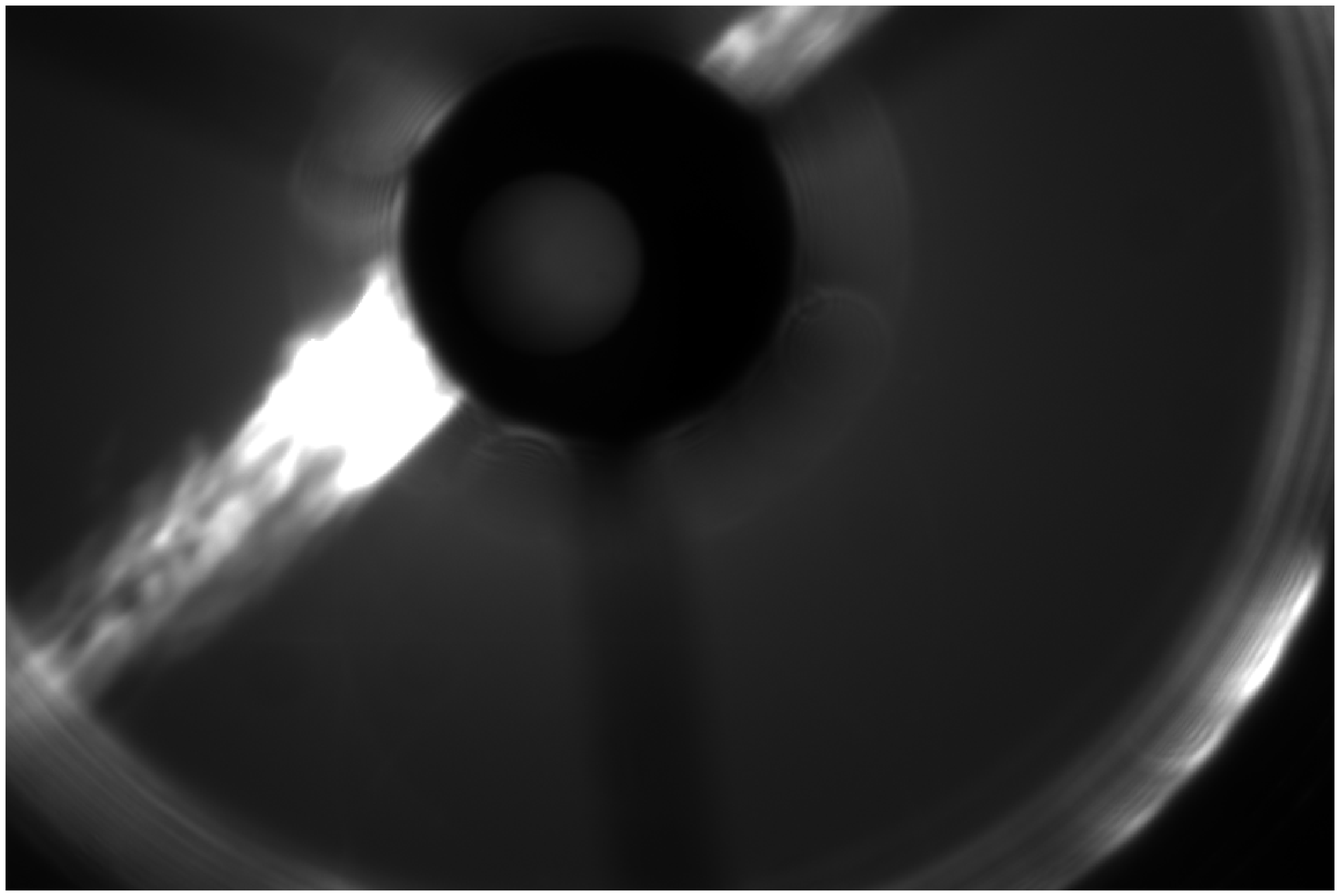}
}
\vspace{-0.33\textwidth}   
\centerline{\Large \bf     
\hspace{0.0 \textwidth}  \color{white}{(a)}
\hspace{0.415\textwidth} \color{white}{(b)}
\hfill}
\vspace{0.29\textwidth}    
\centerline{
\includegraphics[width=0.495\textwidth,clip=]{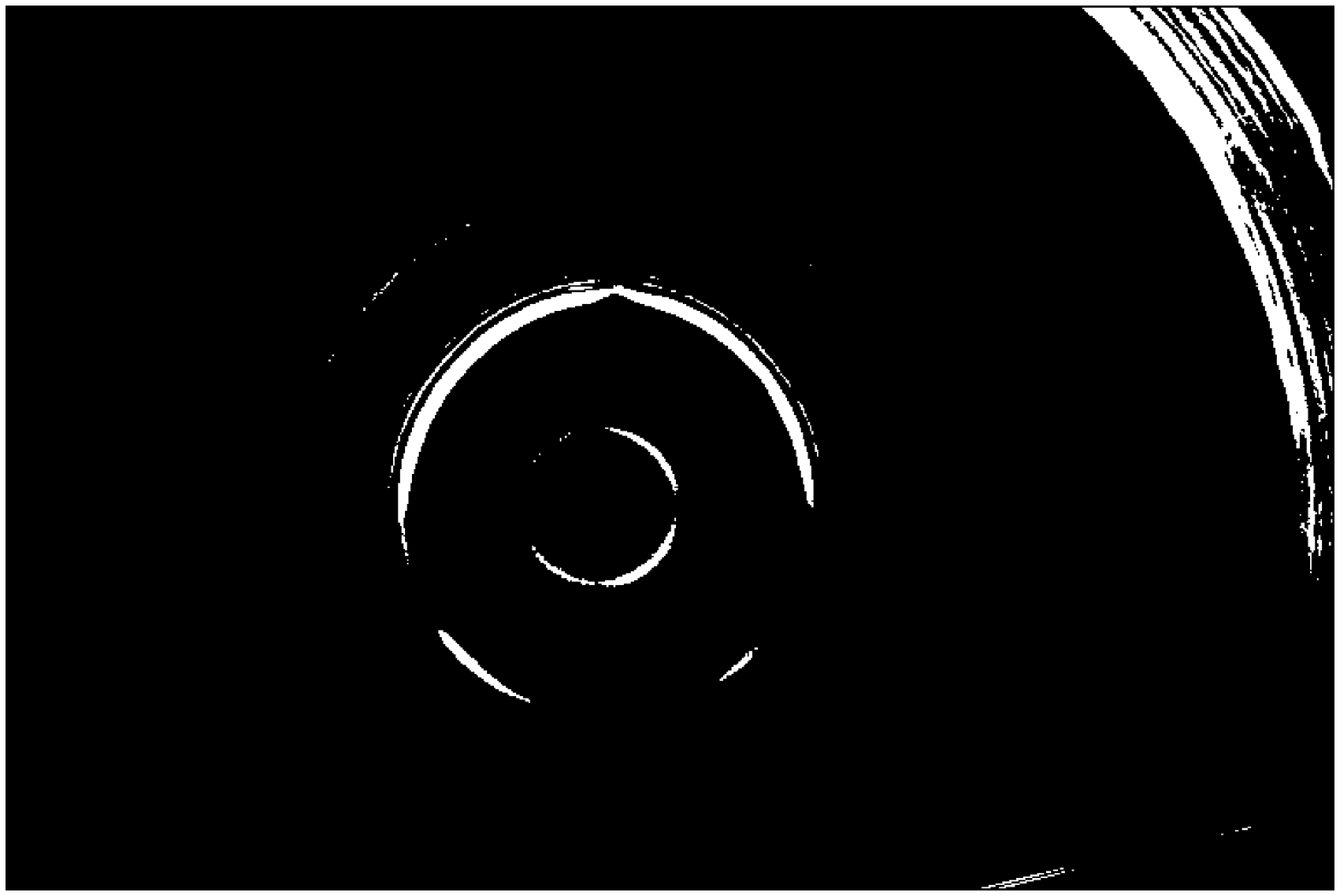}
\includegraphics[width=0.495\textwidth,clip=]{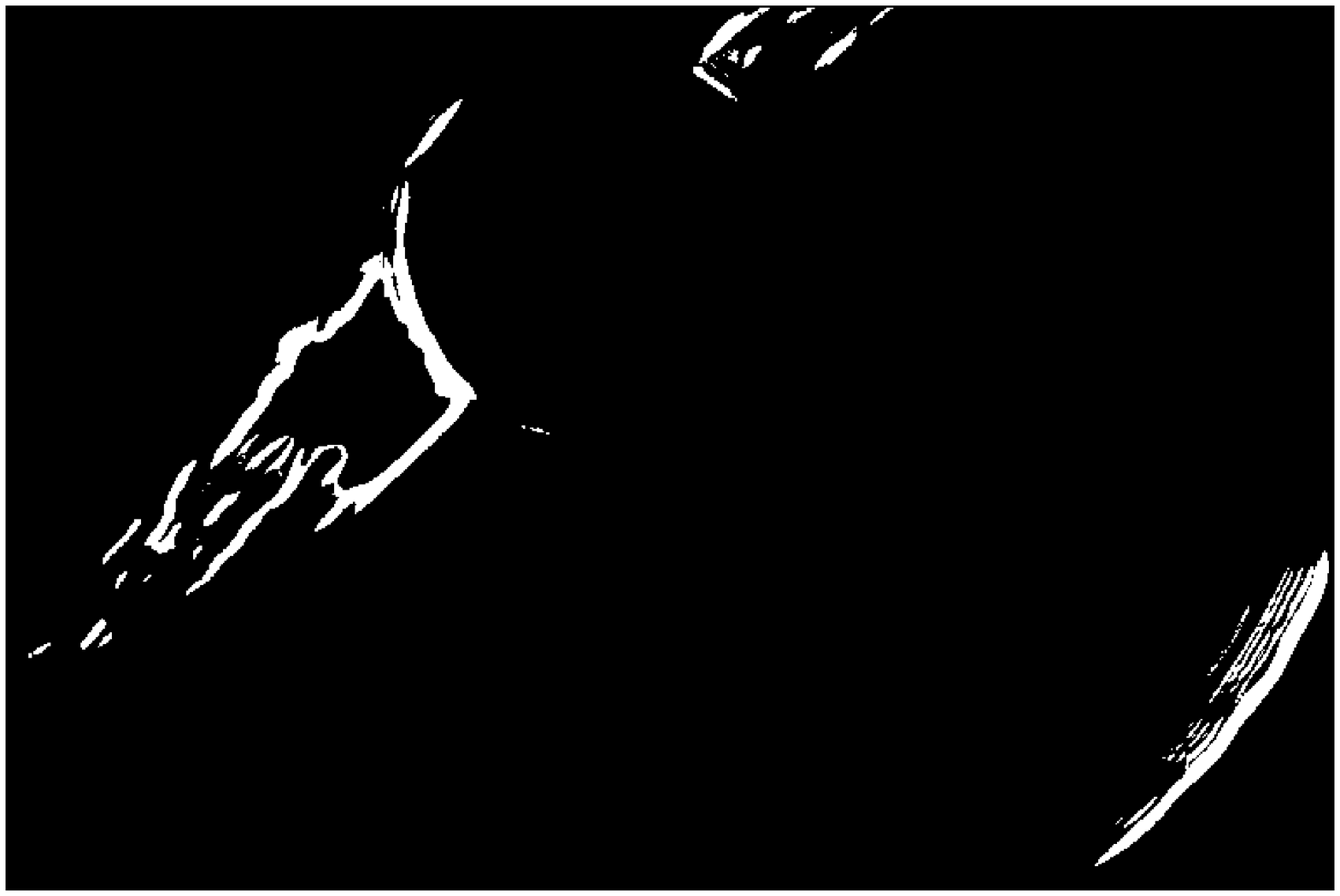}
}
\vspace{-0.33\textwidth}   
\centerline{\Large \bf     
\hspace{0.0 \textwidth}  \color{white}{(c)}
\hspace{0.415\textwidth} \color{white}{(d)}
\hfill}
\vspace{0.29\textwidth}    
\caption{(a): example of normal SBM image. (b): a typical abnormal SBM image corrupted by cloud. (c) and (d): corresponding edge-detecting results obtained by the method of \citet{Zhao2014}.}
\label{fig:CorruptedExp}
\end{figure}

Our SBM is designed independently but has the same function as DKIST SBMs \\   \citep{Liu2012b}. It is based on an externally occulted coronagraph, while the external occulter is replaced with a ND4 filter (a neutral density filter with a nominal optical density of 4). Thus it can simultaneously image the Sun and its nearby sky regions. The details of the instruments are thoroughly described by \citet{Lin2004}. Figure~\ref{fig:CorruptedExp} (a) gives an example of the image obtained by the SBM. The normalized sky brightness with respect to the solar disk center intensity is used to evaluate the coronal observations in optical wavelengths. Therefore, a valid sky region must be established outside of the diffraction rings and between the occulter supporting arms. In \citet{Zhao2014}, we present an automatic method for processing SBM data, which can identify the solar disc and the sky areas around it as well as the excluded part of the supporting arms. This method works well for SBM's data when the day is clear and sunny.

\begin{figure}    
\centerline{
\includegraphics[width=0.495\textwidth,clip=]{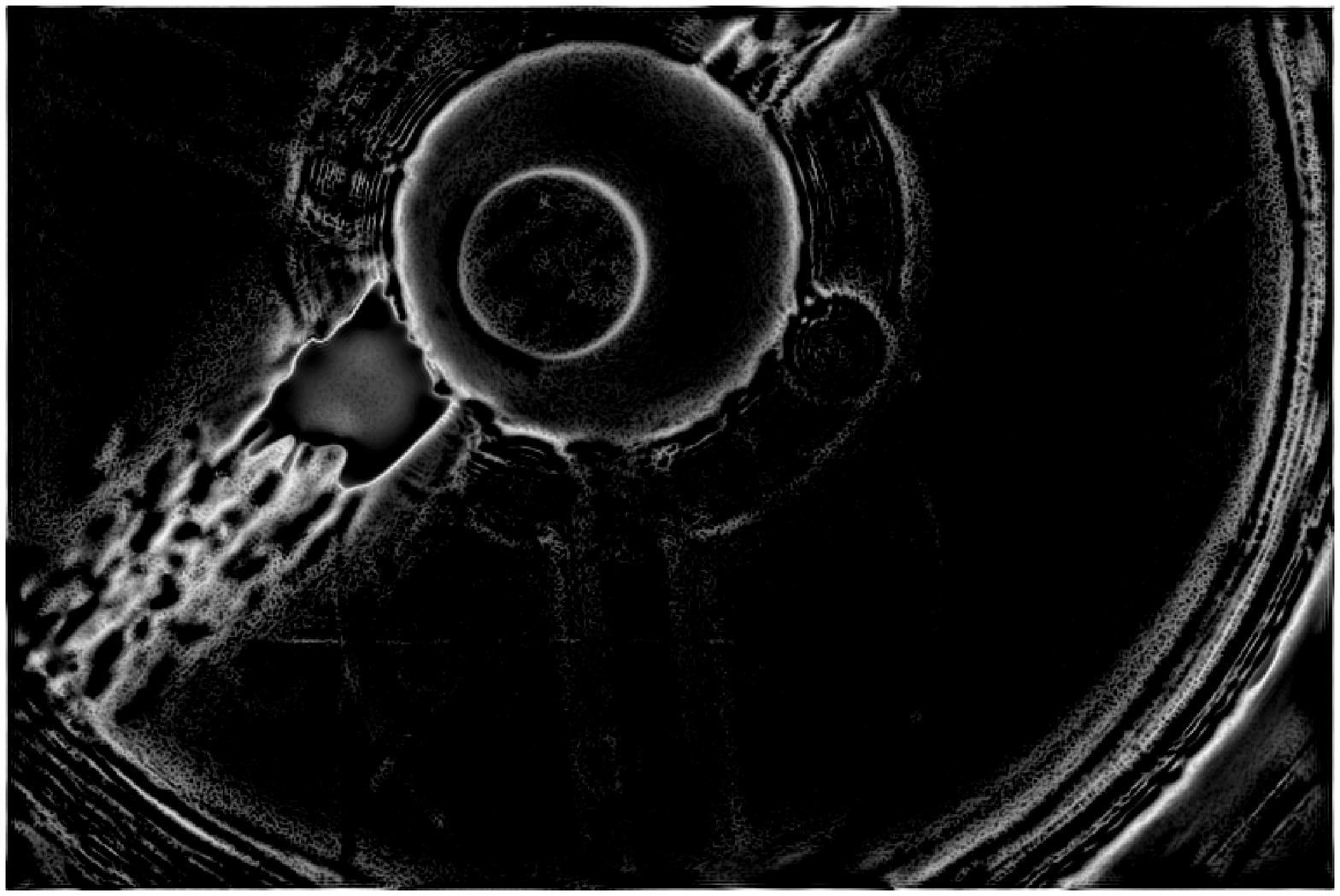}
\includegraphics[width=0.495\textwidth,clip=]{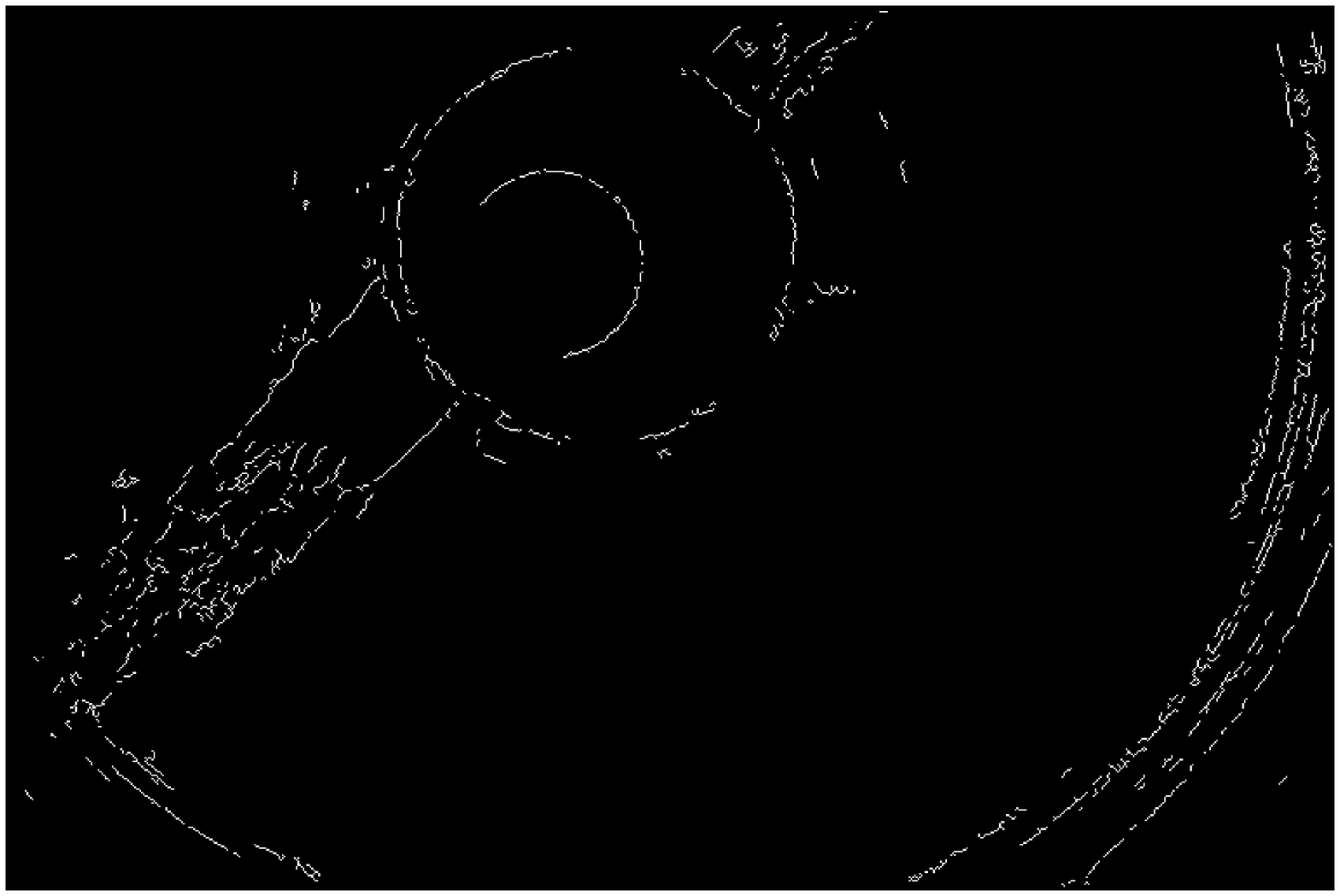}
}
\vspace{-0.33\textwidth}   
\centerline{\Large \bf     
\hspace{0.0 \textwidth}  \color{white}{(a)}
\hspace{0.415\textwidth} \color{white}{(b)}
\hfill}
\vspace{0.29\textwidth}    
\caption{(a): phase congruency calculated by monogenic filters. (b): result of Nonmaxima Suppression of (a).}
\label{fig:CorruptedPC}
\end{figure}

For long-term statistical analysis, it is inevitable that parts of the data are corrupted by cloud. Figure~\ref{fig:CorruptedExp} (b) gives an example of typical abnormal SBM image. The steep gradient of the bright region's edge leads to a falsely identified solar limb, and thus the automation of the data processing is violated. These abnormal data are difficult to exclude in advance when the amount of data is large. In order to overcome this problem, a modern edge-detection algorithm based on the phase congruency principle is adopted to improve the automatic data processing for SBM. The details of this edge detection algorithm are introduced in \autoref{sec:App}. The results of edge detection are shown in Figure~\ref{fig:CorruptedPC}. The limb of the Sun is very clear, because the phase congruency is a dimensionless measure of image features independent of the local contrast. Further steps such as edge extraction and locating the position of supporting arms are still performed following \citet{Zhao2014}. 

The improved algorithm can deal with all of the data taken at Lijiang Observatory in 2011 without changing any parameters. The automatic procedure provides us with a raw temporal sequence of normalized sky brightness. Although the abnormal data will not interrupt the procedure, they must be removed for further analysis. Our temporal sampling interval is about 1.5 minutes; the total amount of data taken at Lijiang Observatory in 2011 is very large. Hence the abnormal data are difficult to exclude manually.

The medians of five-three-Hanning weighted average smoothing is a statistical method of abnormal value removal \citep{Tukey1977,Tukey1981}, which is also termed Tukey's 53H method. According to Tukey's 53H method, the temporal sequence of sky brightness can be treated as a linear combination of a regular function and a fluctuating stochastic process. Because the fluctuations caused by measuring errors should be stationary, the data points with large deviation can be rejected. In detail, let $y_0(t)$ be the raw time sequence of sky brightness, then Tukey's 53H method can be expressed as follows:
\begin{itemize}
\item Constructing a sequence $y_1(t)$ from the median of five data points of $y_0(t)$.
\item Constructing a sequence $y_2(t)$ from the median of three data points of $y_1(t)$.
\item Applying Hanning filter on $y_2(t)$ to obtain the final smoothed curve $y_3(t)$.
\item Calculating the residual $\Delta y=\vert y_0-y_3\vert$ then comparing it with $k\sigma$, where $k$ is a predetermined threshold and $\sigma$ is the standard deviation of $y_0$.
\end{itemize}

\begin{figure}   
\centerline{
\includegraphics[width=0.495\textwidth,clip=]{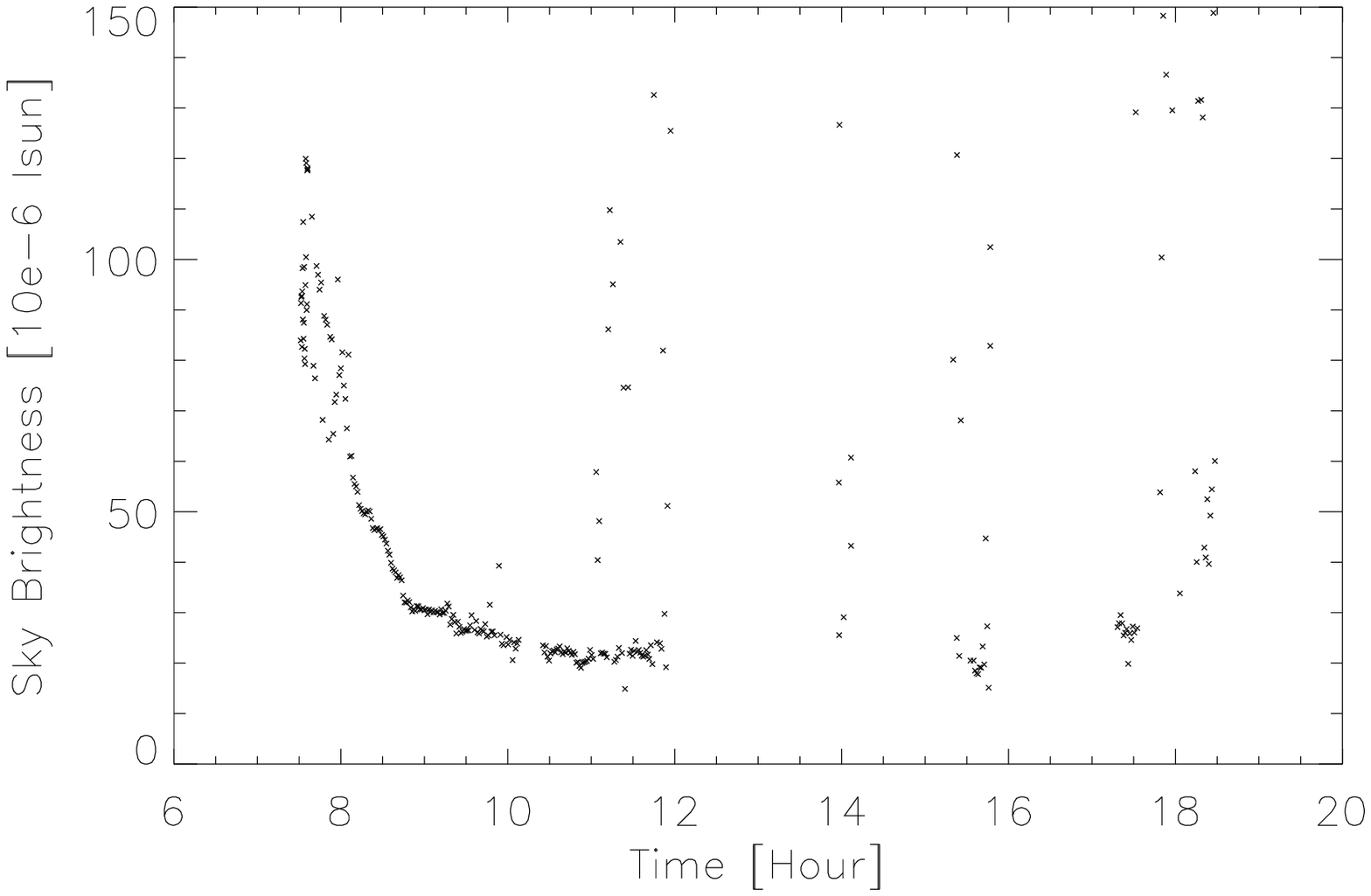}
\includegraphics[width=0.495\textwidth,clip=]{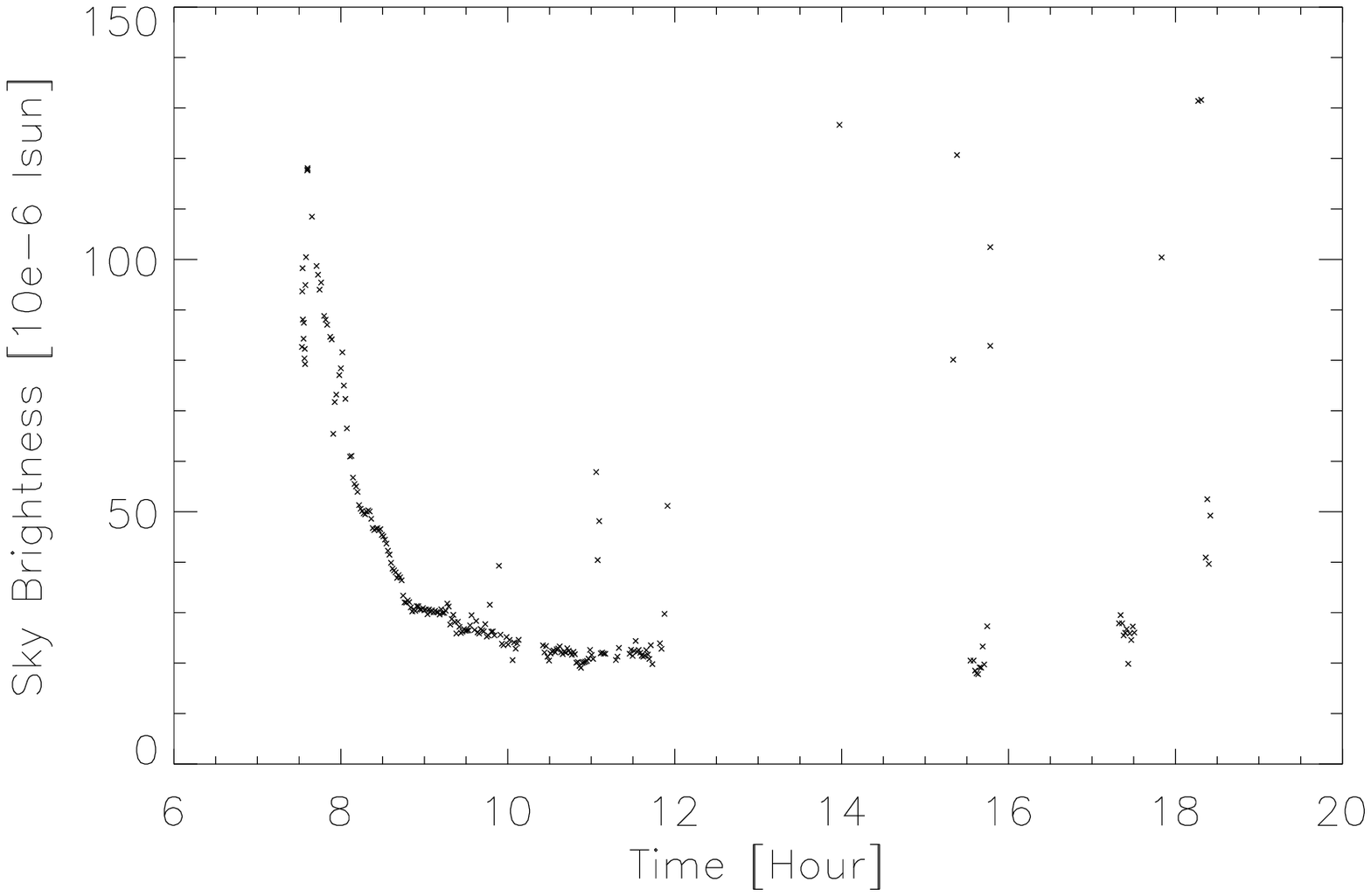}
}
\vspace{0.0\textwidth}   
\centerline{\Large \bf     
\hspace{0.225\textwidth} \color{black}{(a)}
\hspace{0.415\textwidth} \color{black}{(b)}
\hfill}
\vspace{0.0\textwidth}    
\centerline{
\includegraphics[width=0.495\textwidth,clip=]{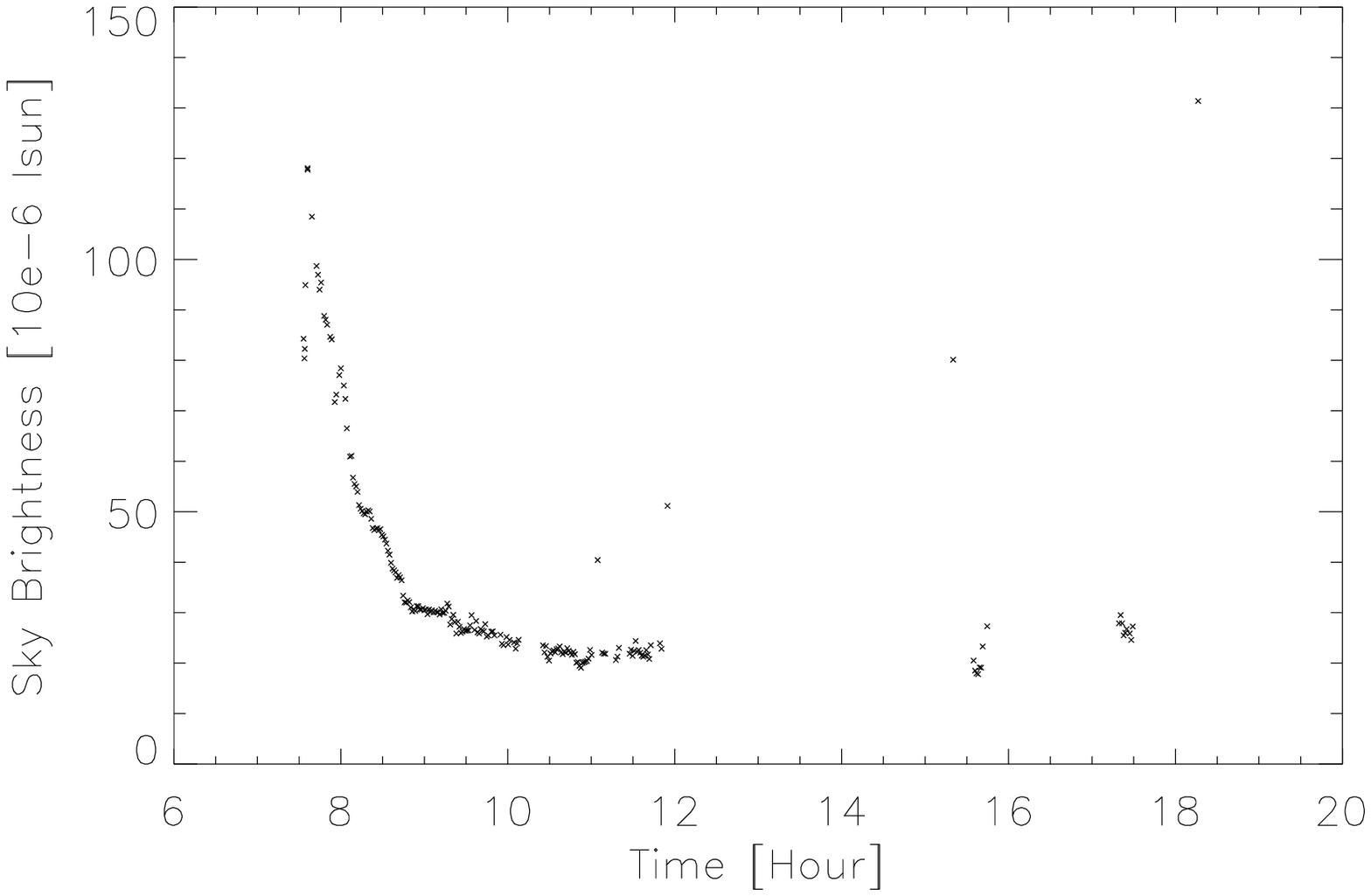}
\includegraphics[width=0.495\textwidth,clip=]{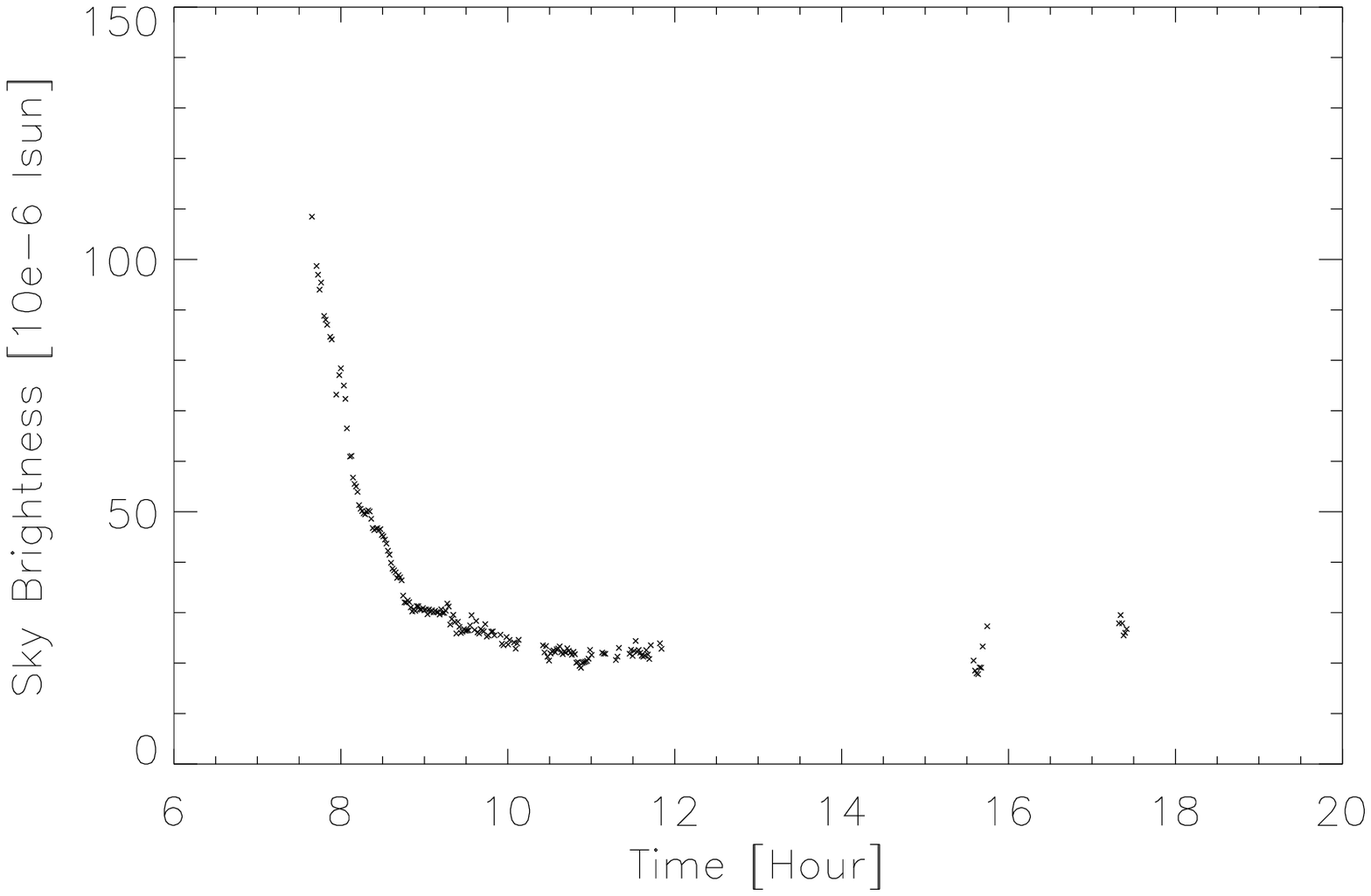}
}
\vspace{-0.0\textwidth}   
\centerline{\Large \bf     
\hspace{0.225\textwidth} \color{black}{(c)}
\hspace{0.415\textwidth} \color{black}{(d)}
\hfill}
\vspace{0.02\textwidth}    
\caption{(a): time sequence of raw SBM data. (b),(c),(d) results of Tukey's 53H method after 1,2,3 recursive iterations, respectively. The predetermined threshold is 0.05}
\label{fig:TK53H}
\end{figure}
Due to the huge amounts of data and discrepancies between days, it is difficult to determine a universal threshold for all of the data. In order to avoid over-rejection, we adopt a small threshold, $k=0.05$, and we apply Tukey's 53H method recursively. Figure~\ref{fig:TK53H} describes the recursive process. Because Tukey's 53H method ignores the first and last four points of the data sequence, the numbers of recursive iterations should not be too large. For the purpose of our present work, choosing three iterations is found to be sufficient.

Tukey's 53H method works well for the data corrupted by cloud. However, there are still some values that are large with respect to the typical varition of the sky brightness after the data selection. Because the SBM is designed to be portable, and in the absence of a guiding system and a high-precision tracking system, some abnormal data are due to the pointing error. These data are usually in a short period and the values of the sky brightness are all large but change slowly and continuously in this period. Therefore, Tukey's 53H method cannot recognize these ''regular'' data based on statistical tendency. We apply a threshold method on the normalized sky brightness at unit airmass as a further constraint. The threshold is 30 ppm/airmass before the local noon and 50 ppm/airmass after the local noon, because the observed value is usually large in the afternoon at Lijiang Observatory. We checked the filtered data and find all of the abnormal data are excluded perfectly. After the data selection, the data are ready for our further statistical analysis.

\section{Sky Brightness Statistics}\label{sec:Stat}
\begin{figure}   
\centering
\includegraphics[width=0.85\textwidth,clip=]{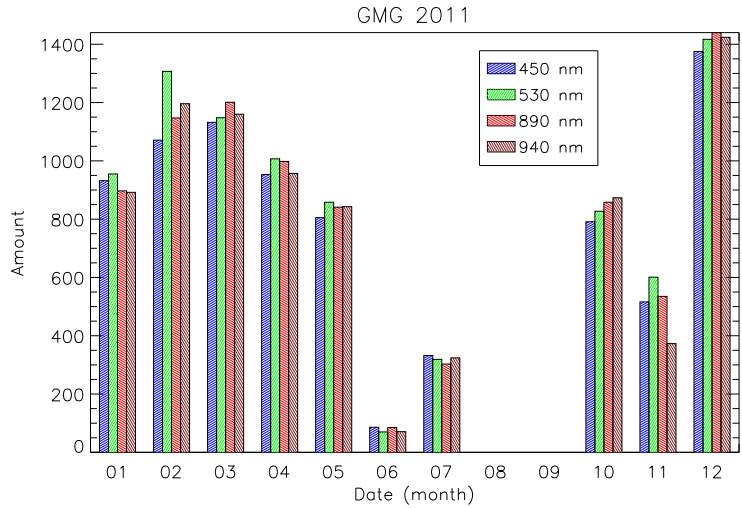}
\caption{The valid sample amounts with month for the sky brightness data taken at Lijiang Observatory in 2011.}
\label{fig:StatSamps}
\end{figure}
In this section, we statistically analyze the sky brightness in 2011 at Lijiang Observatory. The observation site is located near the GaoMeigu (GMG) village ($26^{\circ}42^{'}32^{''}N,100^{\circ}01^{'}52^{''}E$) at an altitude of 3200\,m. The climate of Lijiang presents an obvious monsoonal character. In the rainy season (roughly from June to September), observation is difficult due to the weather, and thus most of the data are acquired between the clouds. The observations usually concentrate in a very short time and most of data are severely affected by clouds. Table~\ref{tab:inSep} gives a preliminary evaluation of the data in the rainy season. We take the data in September as an example and the results show that only about 5\,\% of the observational samples are valid and the measured sky brightnesses are significantly higher than at other times, because these data are mostly acquired between clouds and the high concentration of aerosols gives rise to high sky brightness. Therefore, they are excluded in our further statistics. The numbers of valid samples are shown in Figure~\ref{fig:StatSamps}. 
\begin{table}
\caption{Sky brightness taken at Lijiang Observatory in September, 2011.}
\label{tab:inSep}
\begin{tabular}{cccc}     
\hline                   
Wavelength & Total Sample & Valid Sample & Average Value\\
  ($nm$)   & Numbers      & Numbers      & ($ppm/airmass$)\\  
\hline
450  & 545 & 21 & 72.92 \\
530  & 558 & 23 & 64.88 \\
890  & 573 & 31 & 106.07\\
940  & 533 & 46 & 82.60 \\
\hline
\end{tabular}
\end{table}

The scattered light in the optical system of the SBM is generally measured based on the knowledge of how sky brightness varies as a function of air mass throughout the course of the day \citep{Mar1990,Lin2004,Liu2012b}. This method postulates stable atmospheric conditions, while atmospheric stability deteriorates quickly at the Lijiang site in the afternoon and our dataset does not contain a day with verifiable good sky conditions suitable for instrumental scattered light calibration. As a compromise, we use the sky-brightness values corrected by the instrumental scattered light measured by \citet{Liu2012b} to establish a nominal baseline of the coronal sky quality of the site. \citet{Liu2012b} use the same SBM but with a different ND4 filter and the scattered light are 0.77, 0.81, 2.12 and 3.26 millionths for the  450\,nm, 530\,nm, 890\,nm and 940\,nm respectively.

\begin{figure}   
\centerline{
\includegraphics[width=0.495\textwidth,clip=]{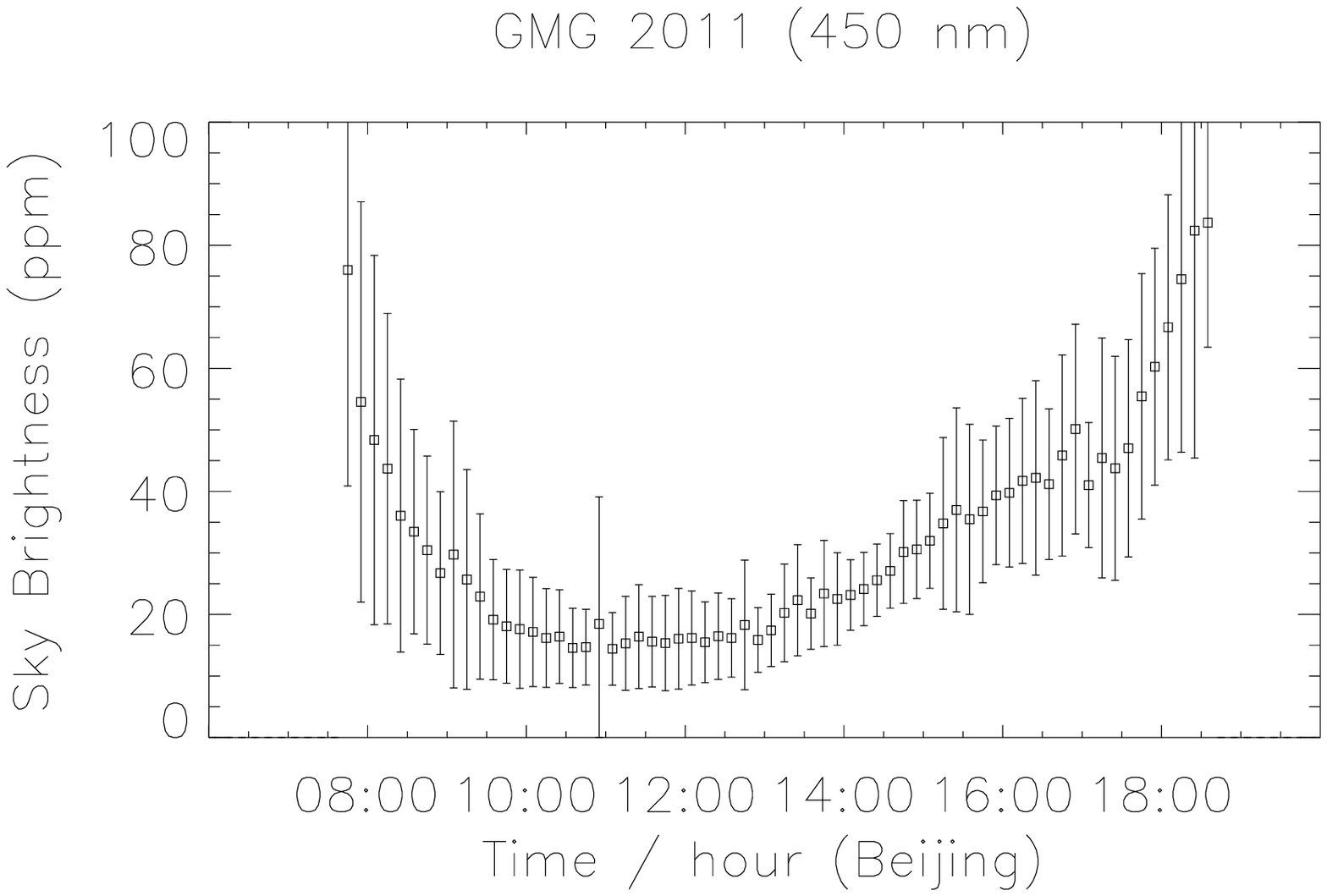}
\includegraphics[width=0.495\textwidth,clip=]{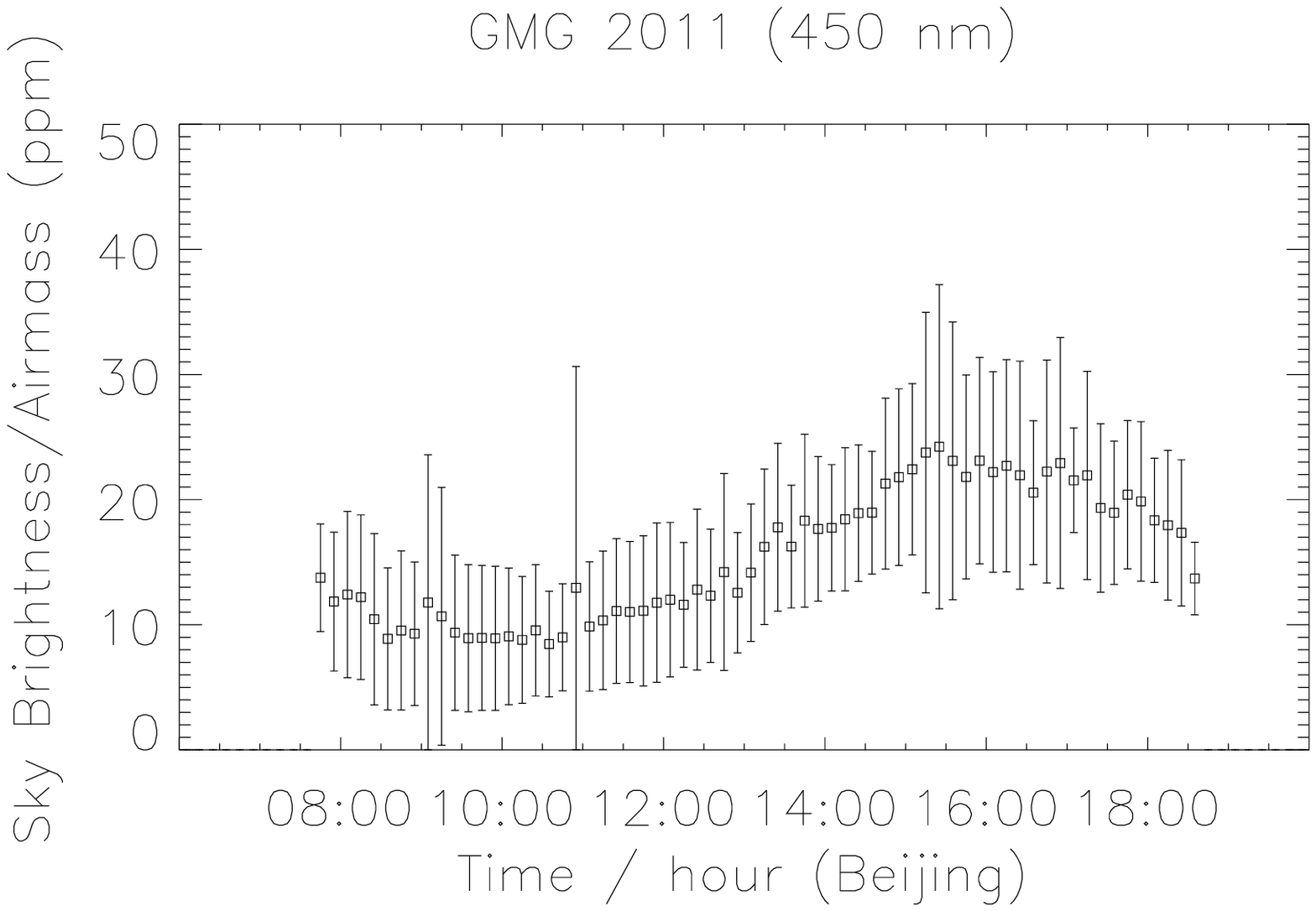}
}
\vspace{-0.28\textwidth}   
\centerline{\Large \bf     
\hspace{0.225\textwidth} \color{black}{(a)}
\hspace{0.415\textwidth} \color{black}{(b)}
\hfill}
\vspace{0.22\textwidth}    
\centerline{
\includegraphics[width=0.495\textwidth,clip=]{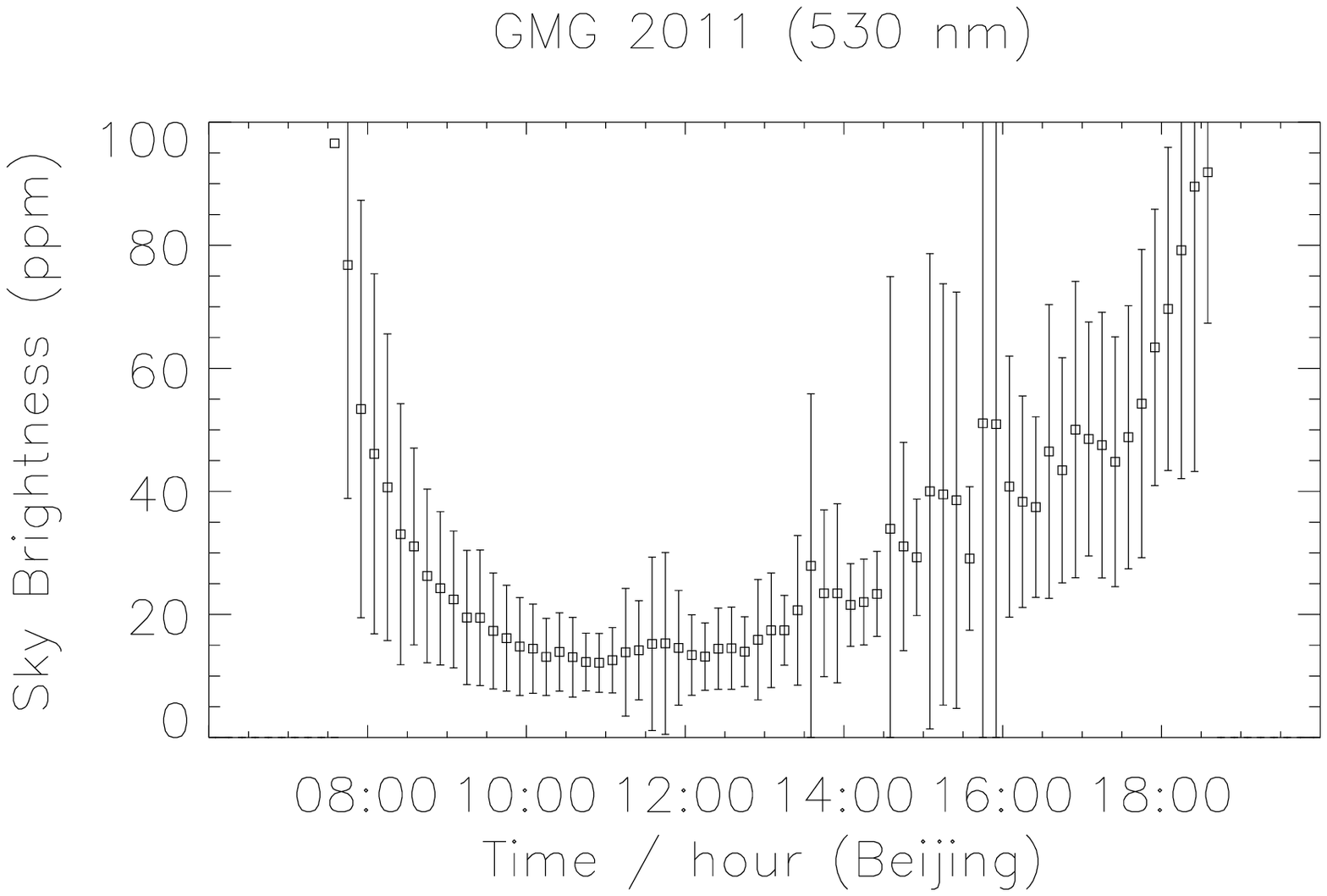}
\includegraphics[width=0.495\textwidth,clip=]{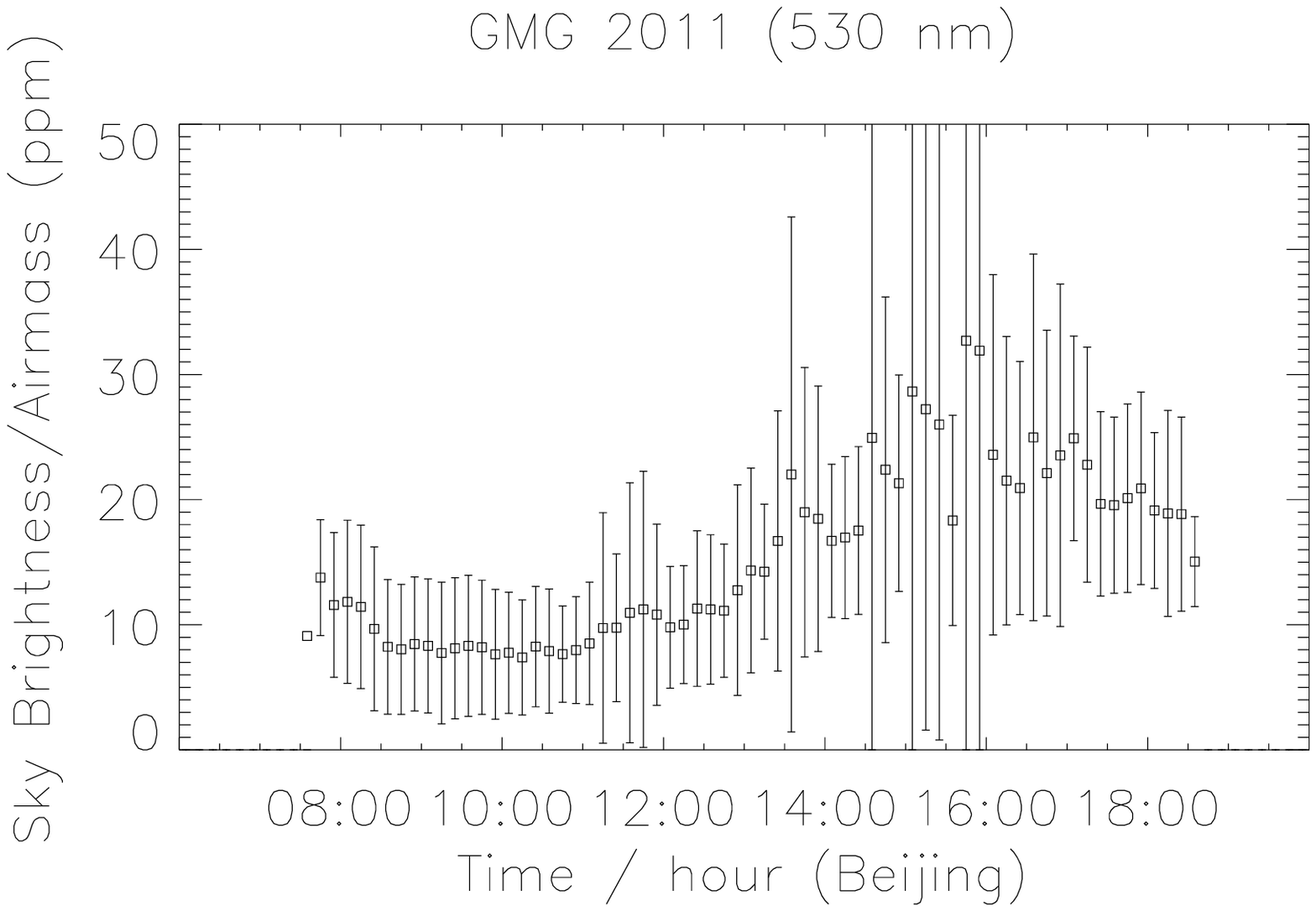}
}
\vspace{-0.28\textwidth}   
\centerline{\Large \bf     
\hspace{0.225\textwidth} \color{black}{(c)}
\hspace{0.415\textwidth} \color{black}{(d)}
\hfill}
\vspace{0.22\textwidth}    
\centerline{
\includegraphics[width=0.495\textwidth,clip=]{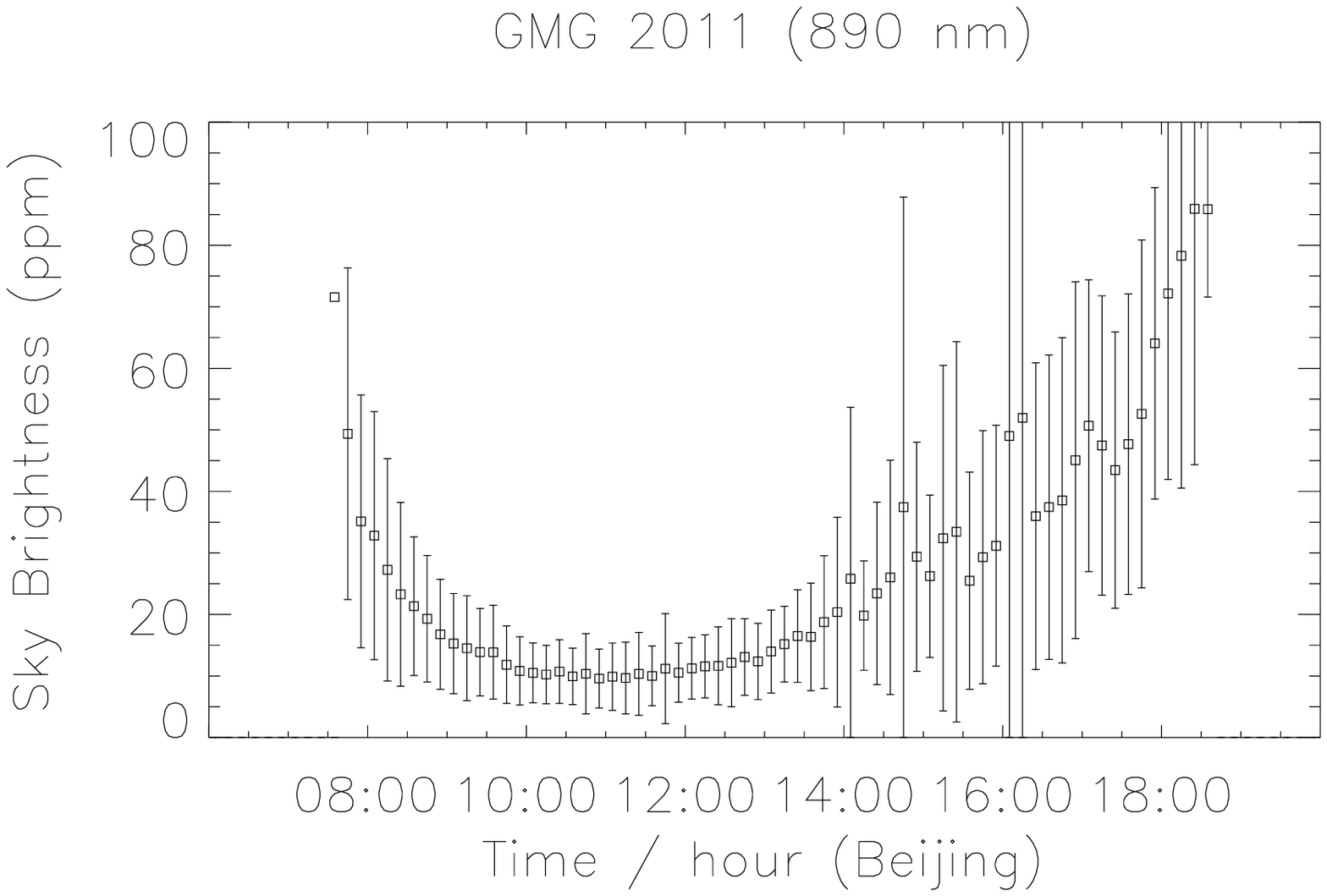}
\includegraphics[width=0.495\textwidth,clip=]{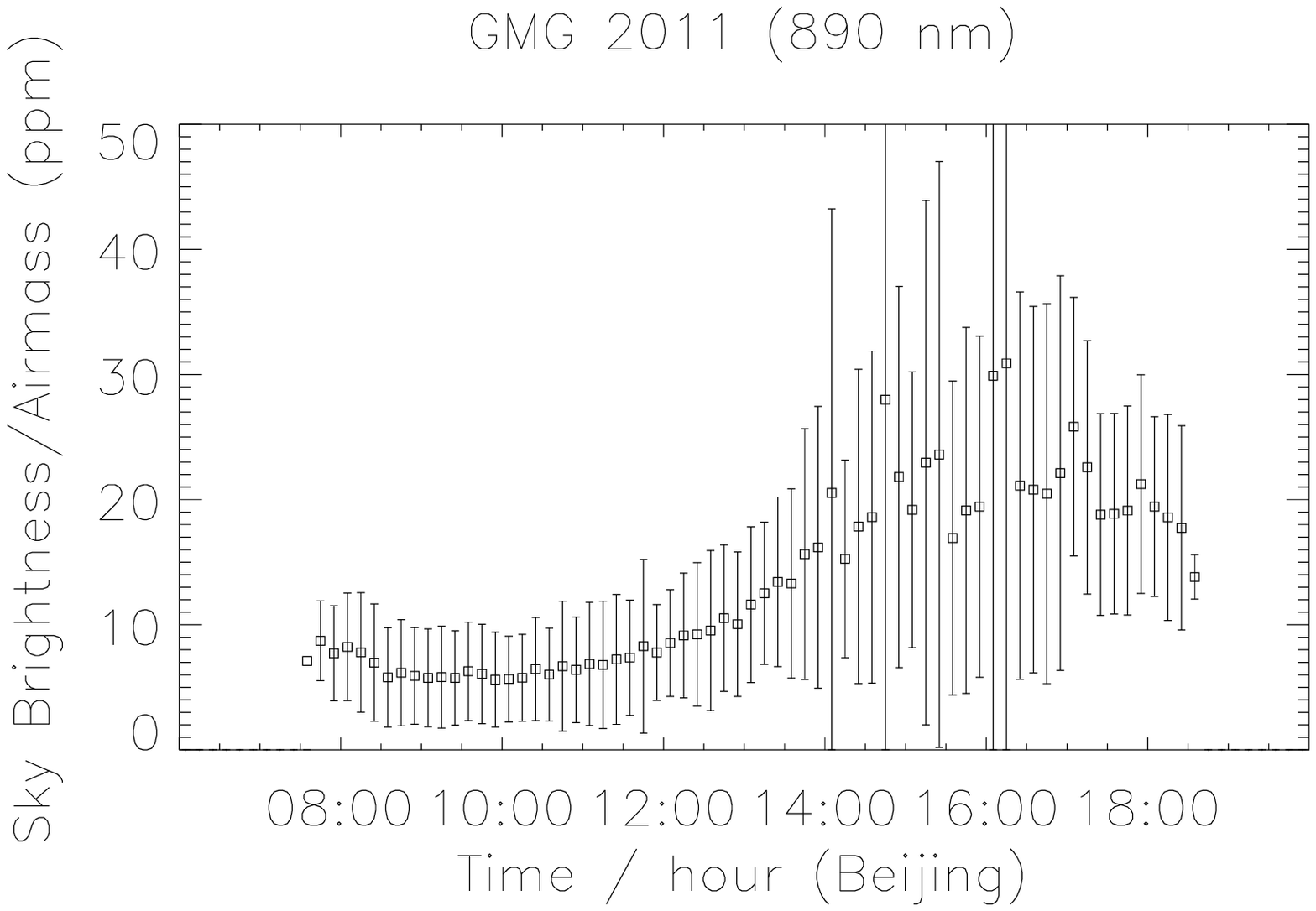}
}
\vspace{-0.28\textwidth}   
\centerline{\Large \bf     
\hspace{0.225\textwidth} \color{black}{(e)}
\hspace{0.415\textwidth} \color{black}{(f)}
\hfill}
\vspace{0.22\textwidth}    
\centerline{
\includegraphics[width=0.495\textwidth,clip=]{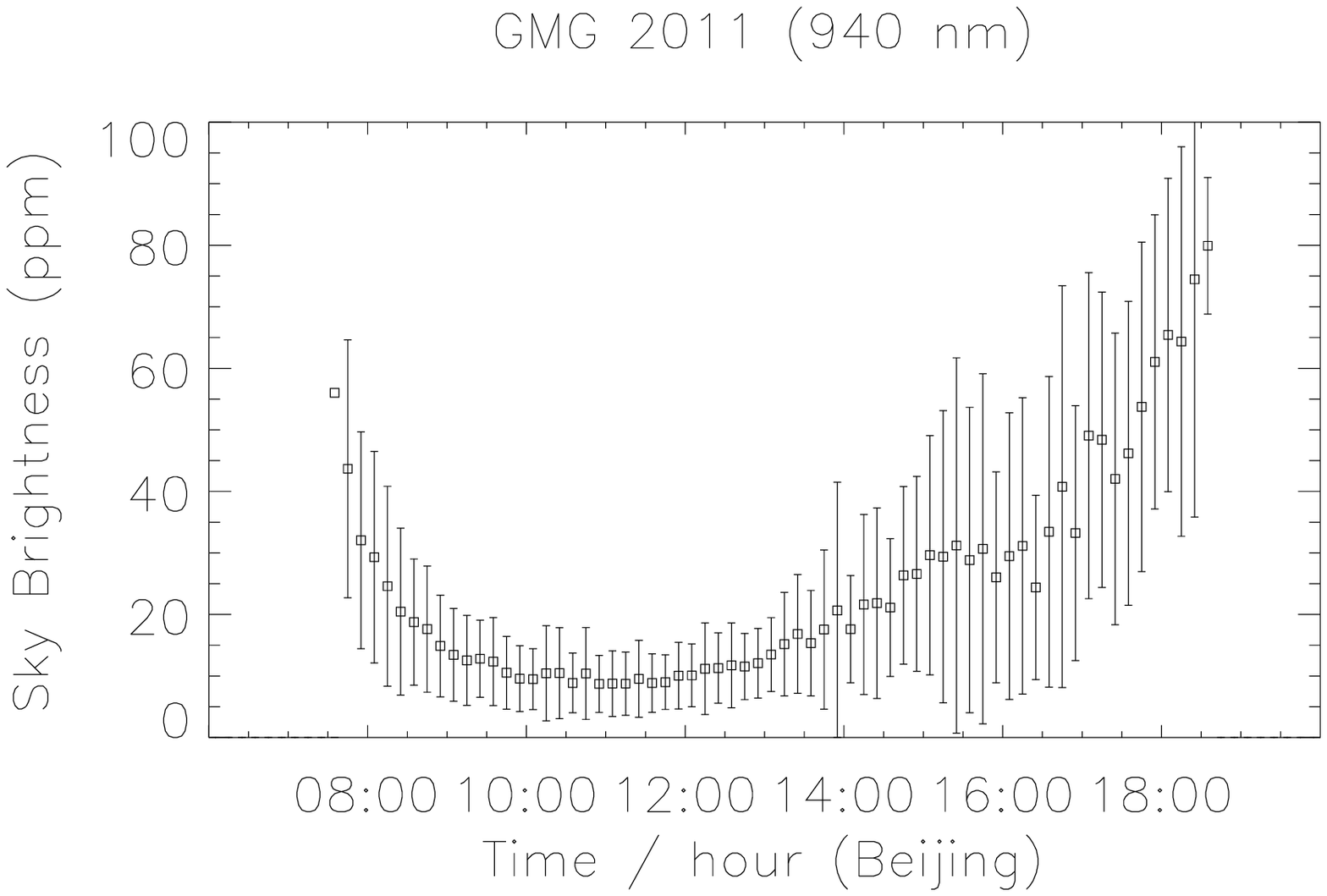}
\includegraphics[width=0.495\textwidth,clip=]{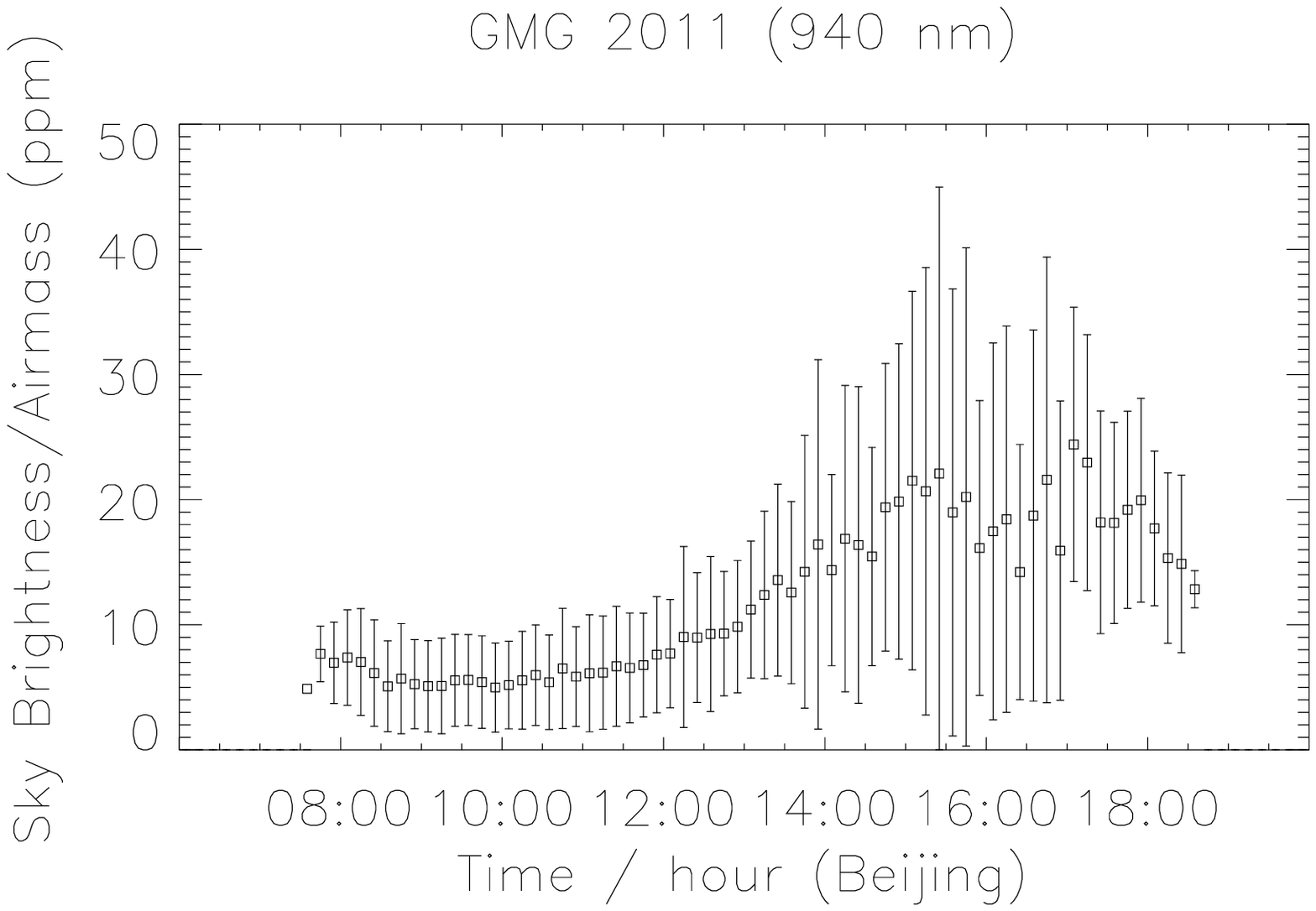}
}
\vspace{-0.28\textwidth}   
\centerline{\Large \bf     
\hspace{0.225\textwidth} \color{black}{(g)}
\hspace{0.415\textwidth} \color{black}{(h)}
\hfill}
\vspace{0.22\textwidth}    
\caption{The diurnal distributions of mean sky brightness at Lijiang Observatory in 2011. The error-bars represent the standard deviations. The left column and right column correspond to the data before and after normalized to zenith respectively}
\label{fig:StatPerUnit}
\end{figure}

The data are gathered at Lijiang Observatory in 2011, sampling from one to ten days per month, mainly depending on the weather conditions. In order to give the diurnal variation, we divide a day into several time intervals of ten minutes and then calculate the mean value and standard deviation in each interval. The observed sky brightness is expected to vary with airmass, which is different at the same time interval on different days. It is necessary to normalize the sky brightness at unit airmass. Both the original and normalized sky brightnesses are calculated and the results are shown in Figure~\ref{fig:StatPerUnit}. On the whole, the normalized sky brightness are mostly under 20\,ppm/airmass and it is not constant. In the morning, the sky brightness changes a little because the sky conditions are usually stable. In the afternoon, it increases and reaches its maximum about one to two hours after the local noon. This tendency is similar to the results of \citet{Labonte2003}, which indicates that it results from the buildup of aerosols. 
\begin{figure}   
\centerline{
\includegraphics[width=0.495\textwidth,clip=]{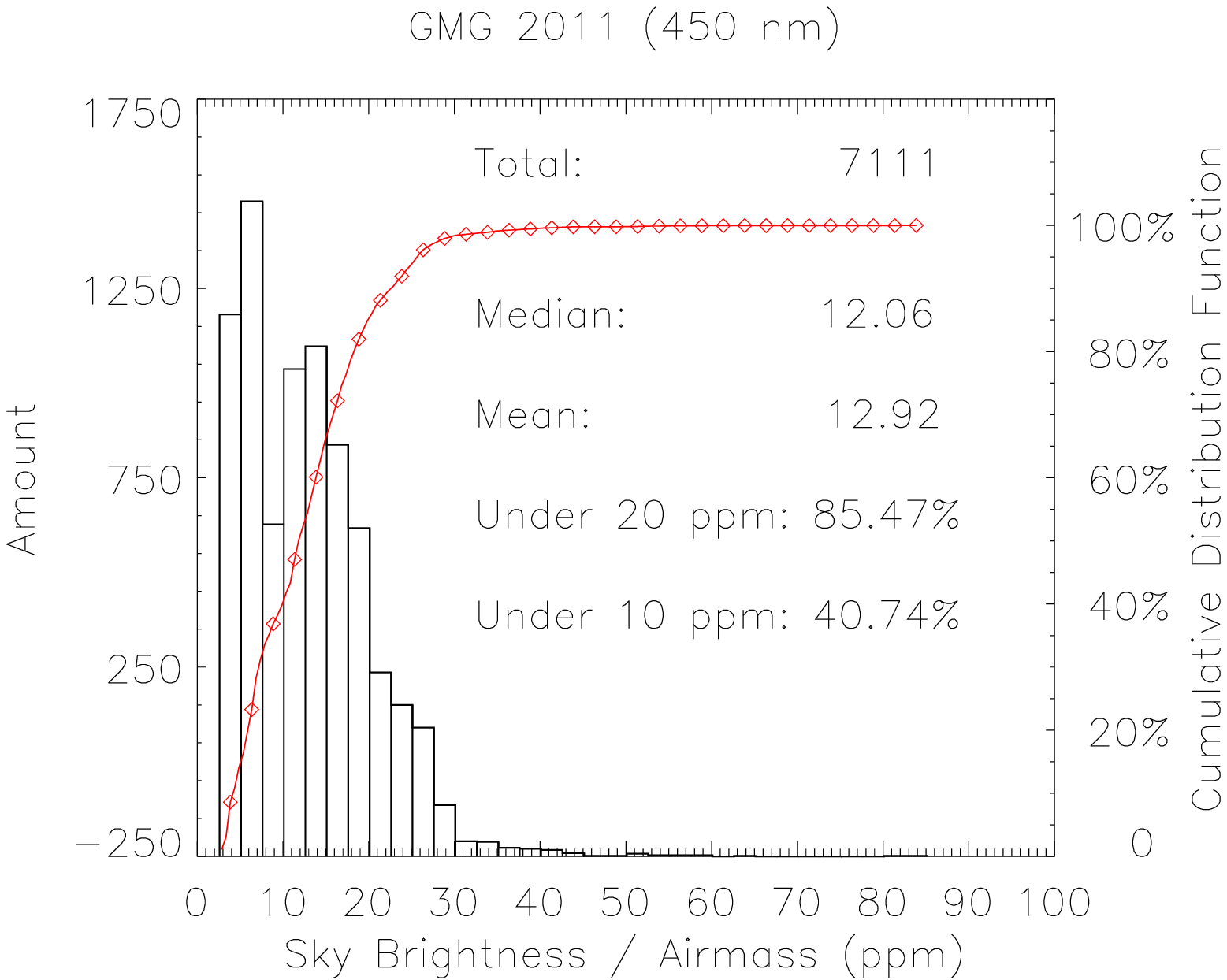}
\includegraphics[width=0.495\textwidth,clip=]{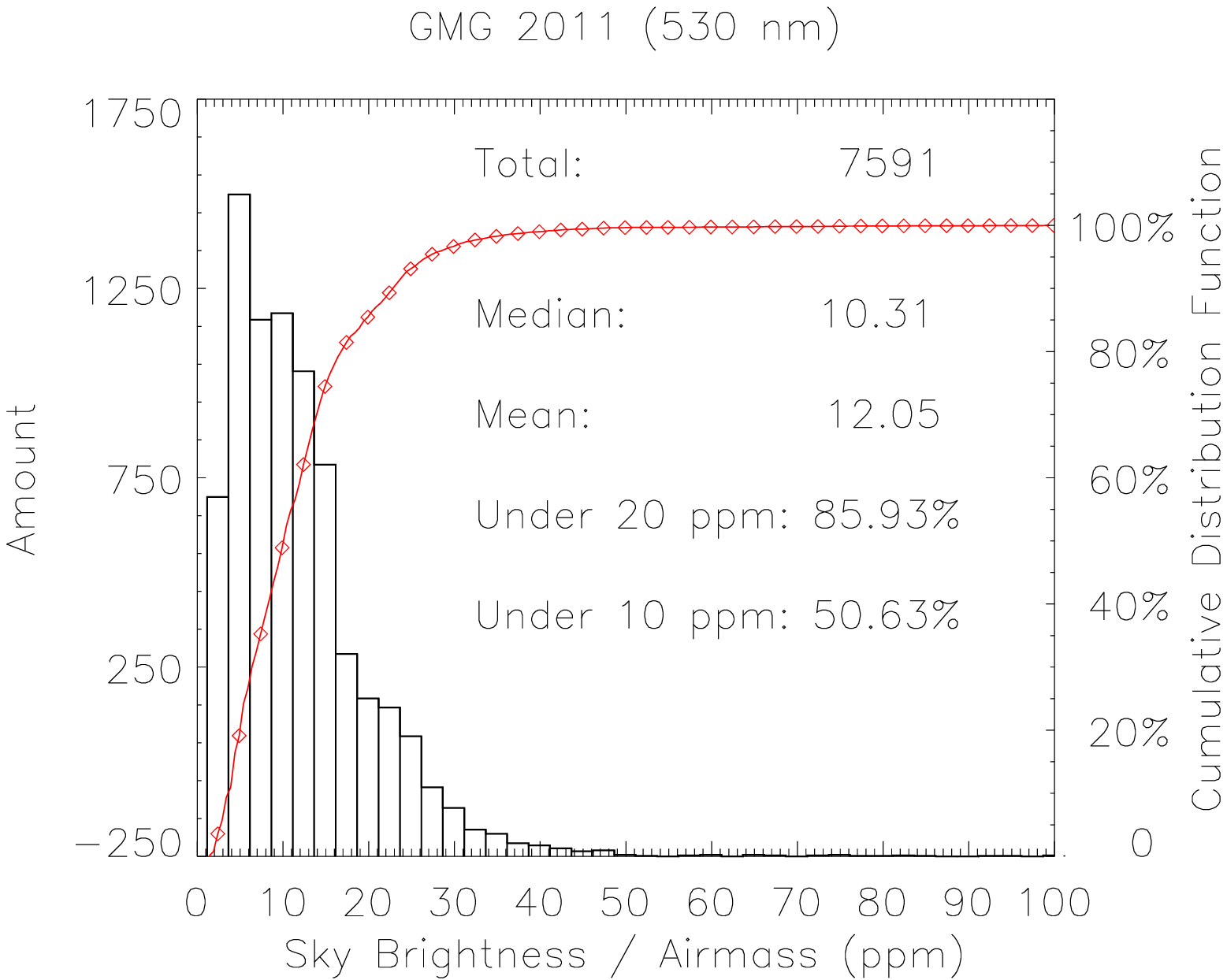}
}
\centerline{
\includegraphics[width=0.495\textwidth,clip=]{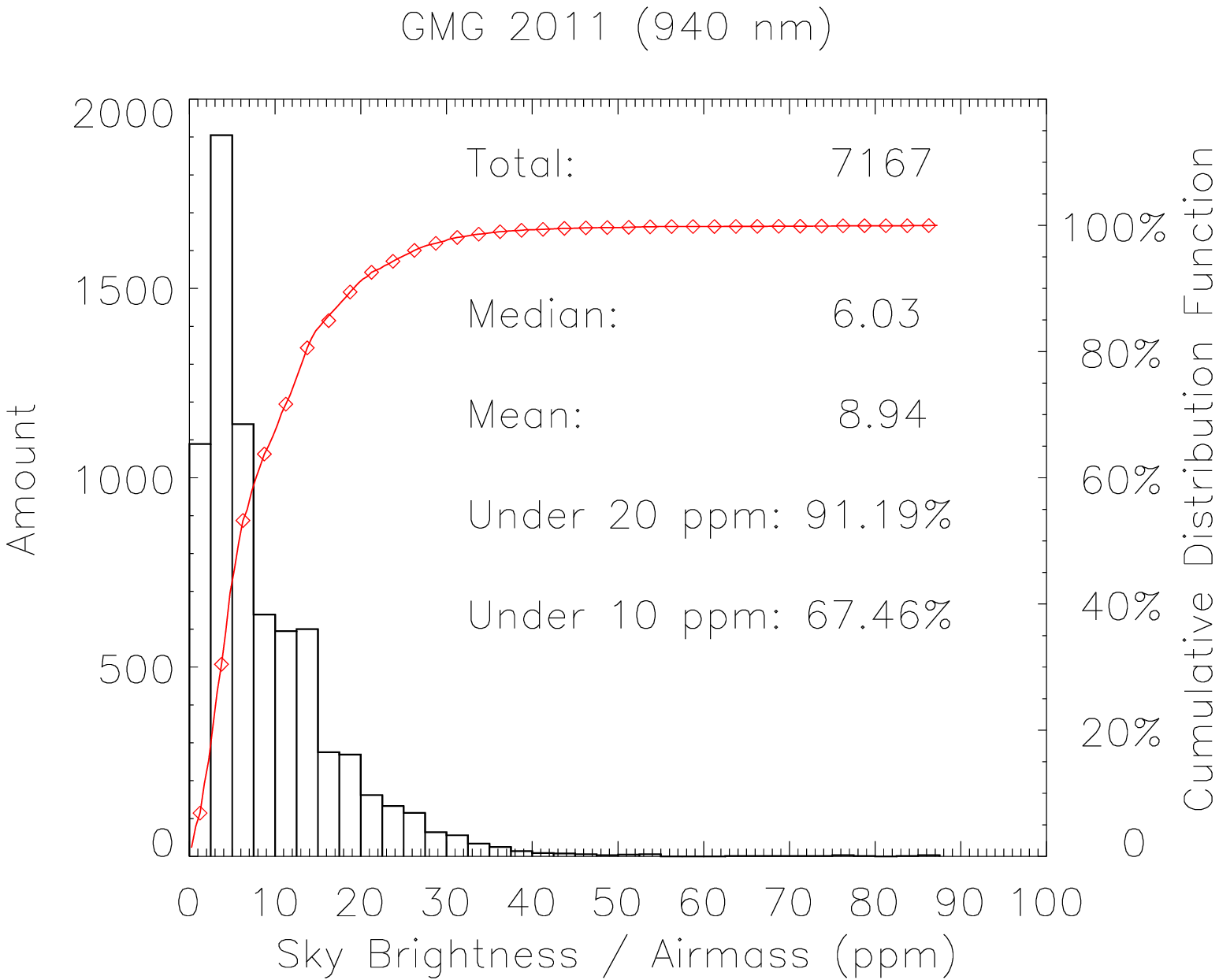}
\includegraphics[width=0.495\textwidth,clip=]{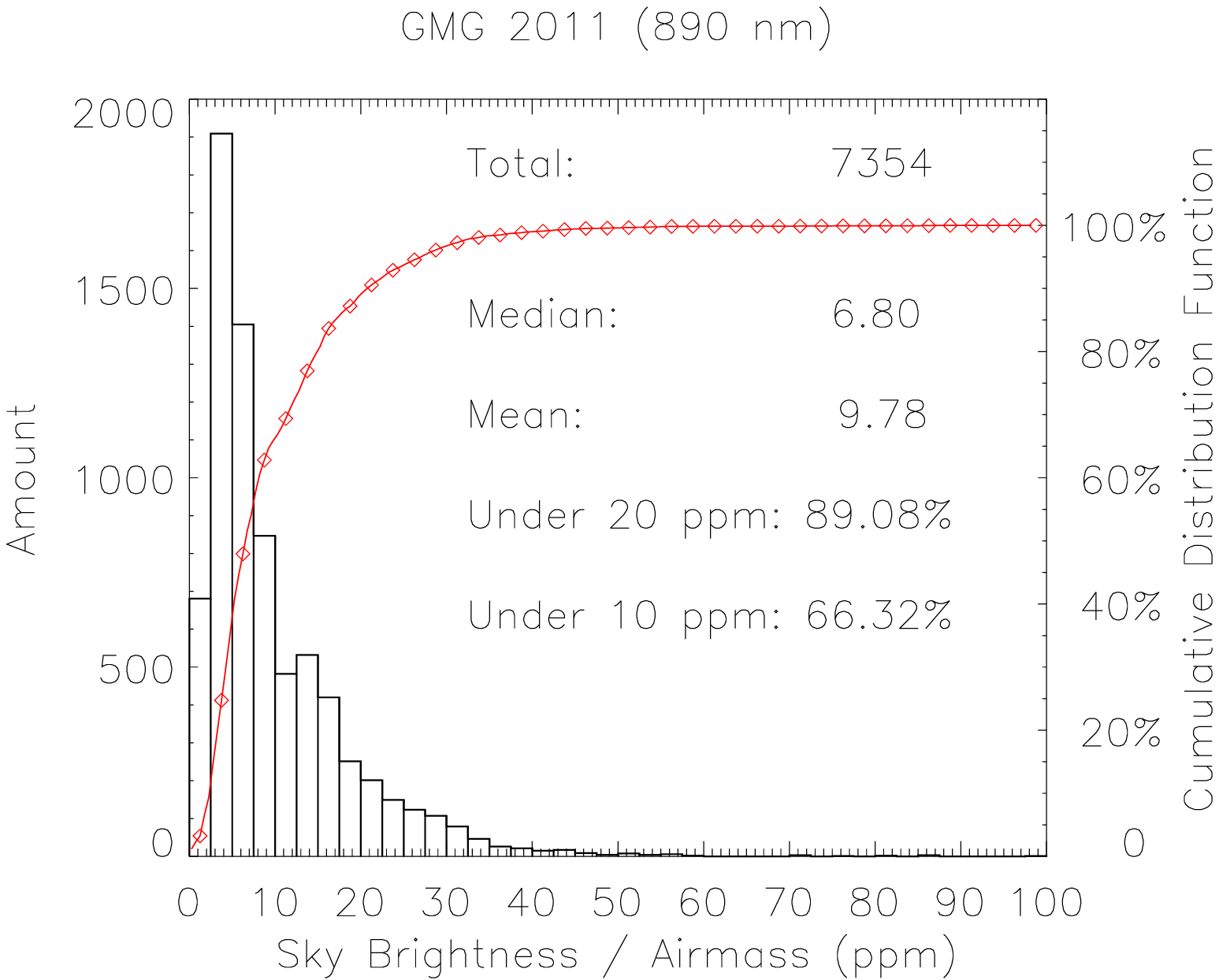}
}
\caption{The probability functions and cumulative distribution functions of the sky brightness based on the valid numbers of our observation at Lijiang Observatory in 2011, which are calculated with bins in steps of 2.5 millionths per airmass.}
\label{fig:StatProb}
\end{figure}

The statistical distribution of the normalized sky brightness and the corresponding parameters are shown in Figure~\ref{fig:StatProb}. The statistics are based on the valid numbers of our observation. The probability density and cumulative distribution functions are calculated with bins in steps of 2.5 millionths per airmass.  For each wavelength, the number of samples is more than 7000 per year. The mean values and standard deviations over the year are $12.92\pm7.39$, $12.05\pm8.50$, $9.78\pm8.68$, and $8.94\pm8.22$ ppm/airmass for 450\,nm, 530\, nm, 890\,nm, and 940\,nm respectively. Both the mean and median value over the year decrease with increasing wavelength.
At an excellent coronagraph site, the sky brightness at noon time is usually around 10 millionths of the solar disk center intensity\citep{Lin2004,Penn2004a,Penn2004b}, so we give the cumulative probabilities for 10\,ppm and 20\,ppm at each wavelength. The results imply very good coronal observation conditions at Lijiang Observatory.

\begin{figure}    
\centerline{
\includegraphics[width=\textwidth,clip=]{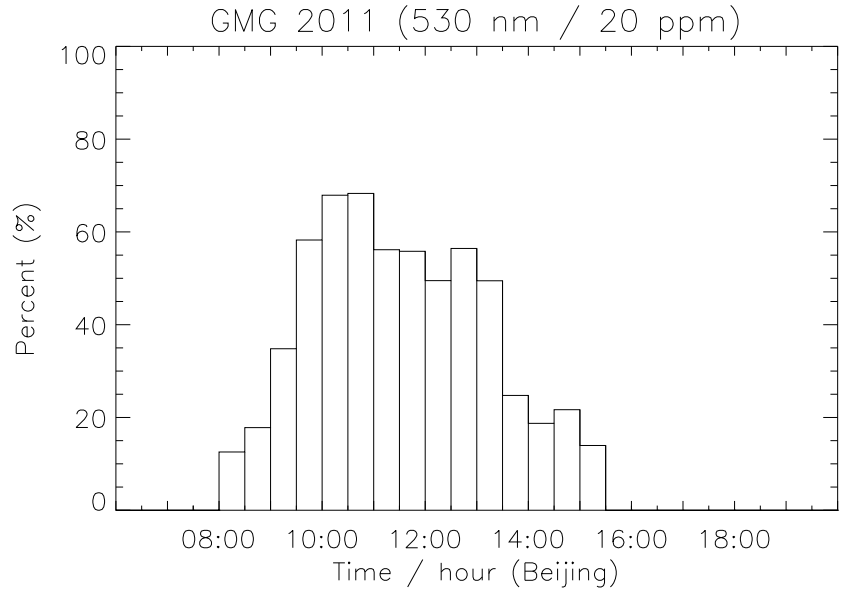}
}
\caption{Time periods that are suitable for coronal observation}
\label{fig:Suitable}
\end{figure}

The 10-cm coronagraph NOGIS (NOrikura Green-line Imaging System), \\which has been working at Lijiang Observatory from the end of 2013, observes the coronal emission line of Fe XIV 5303 \AA. Hence the observation of SBM at 530\,nm (10\,nm bandpass) is of great importance. Based on the local meterological records, the yearly amount of sunshine hours is more then 2000 hours at Lijiang \citep{Wan2009}. The number of meteorological clear days (over 24 hours, total cloud covor equals zero) is 80.5 per year during 1994 to 1996 and it is 84.8 per year averaged from 1959 to 1994 \citep{Zhang1996,Tan2002}. However, our data sampling does not cover every day of the year. In order to estimate which times of day are suitable for coronal observation at Lijiang Observatory, we divide a day into several time intervals of 0.5 hours and then calculate the statistics of sky brightness at actual airmass in each interval. Because our SBM acquires a set of data in a little more than one minute, we consider that it is suitable for coronal observation only if there are more than 20 samples per half hour. Then we calculate the number of samples in each time interval that are under 20 millionths. We normalized the results by the total sample number and obtain a relative evaluation of observing periods for good coronal observation conditions. Figure~\ref{fig:Suitable} demonstrates the probability of conditions suitable for coronal observation on a clear day. On average, sky brightness under 20 millionths accounts for 40.41\,\% of the total observing time, and the most suitble period for coronal observation is 9:00 to 13:00 (Beijing time) at Lijiang Observatory.

\section{Conclusions} \label{sec:con}
We modify the automatic proceeding algorithm proposed by \citet{Zhao2014} and apply it on the data acquired at Lijiang Observatory in 2011. A statistical method of abnormal value removing (Tukey's 53H method) is employed recursively in the measured sky brightness time sequence. It is effective to filter abnormal data avoiding to evaluate huge number of images taken under intricate conditions. 

The sky brightness statistics at Lijiang Observatory in 2011 indicate that the aerosol scattering is of great importance for the diurnal variation. The sky brightness is usually much lower in the morning than that in the afternoon. Generally, good coronal observation conditions can be guaranteed in the morning, while conditions deteriorate quickly in the afternoon.

The statistical median and mean values combined with the cumulative probability imply most of the sky brightness measurements are under 20 millionths per airmass, when the sky is clear and sunny. The yearly amount of sunshine hours is more then 2000 hours, according to local meterological records, and sky brightness under 20 millionths accounts for 40.41 \% of the total observing time in a clear day based on our data samples. The best period of time for coronal observation is from 9:00 to 13:00 (Beijing time).

\section*{Disclosure of Potential Conflicts of Interest}
The authors declare that they have no conflicts of interest.

\begin{acks}
This work was supported by grants from NSFC (11078004, 11533009, 11503084,11603074), and from the Key Laboratory of Geospace Environment, CAS, University of Science \& Technology of China, and from Huairou
Solar Observing Station, National Astronomical Observatories Chinese Academy of Sciences. We are grateful to Prof. Peter Kovesi for his codes of the phase congruency algorithm \citep{KovesiCode}. 
\end{acks}
%

\appendix   
\section{Edge-Detection Algorithm}
\label{sec:App}
\begin{figure}    
\centerline{
\includegraphics[width=0.75\textwidth,clip=]{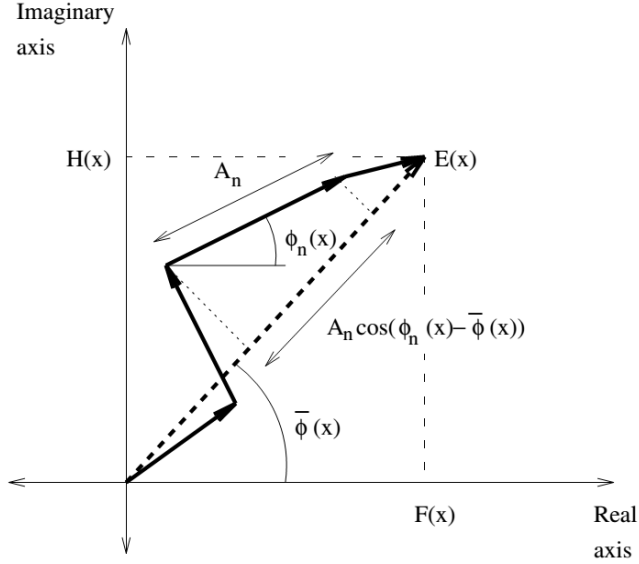}
}
\caption{Polar diagram showing the Fourier components at a location [$x$] in the signal [$F(x)$] plotted head to tail. This figure is adapted from \citet{Kovesi1999}}
\label{fig:PCsch}
\end{figure}

We adopt a method based on phase congruency principle to detect the edges of SBM's images, which can be traced back to the experiment of \citet{Oppenheim1981}. This experiment shows that the amplitude spectrum of an image can be modified considerably, even swapped with that from another image, and the features from the original image will still be seen clearly as long as the phase information is preserved. \citet{Morrone1986} and \citet{Morrone1987} develop a local energy model for feature detection via phase congruency. This model assumes that the compressed image format should be high in information and low in redundancy. It searches for patterns of order in the phase component of the Fourier transform. A phase congruency function [$PC(x)$] at each point $x$ in the signal can be defined as
\begin{equation}
\label{eq:PC}
PC(x) = \max_{\overline{\phi} \in [0,2\pi)}\frac{\sum_n A_n\cos(\phi_n(x)-\overline{\phi}(x))}{\sum_n A_n}
\end{equation}
where $A_n$ represents the amplitude of the \textit{n}th Fourier component, and $\phi_n(x)$ represents the local phase of the Fourier component at position $x$. The relationship between phase congruency, local energy [$E(x)$], and the sum of the Fourier amplitudes can be seen geometrically in Figure~\ref{fig:PCsch}. $PC(x)$ is a measure of the variance of the phase values of the signal, which takes on values between zero and unity, and it provides an illumination and spatial-magnification invariant measure of feature significance \citep{Kovesi1995}.

\begin{figure}   
\centerline{
\includegraphics[width=0.495\textwidth,clip=]{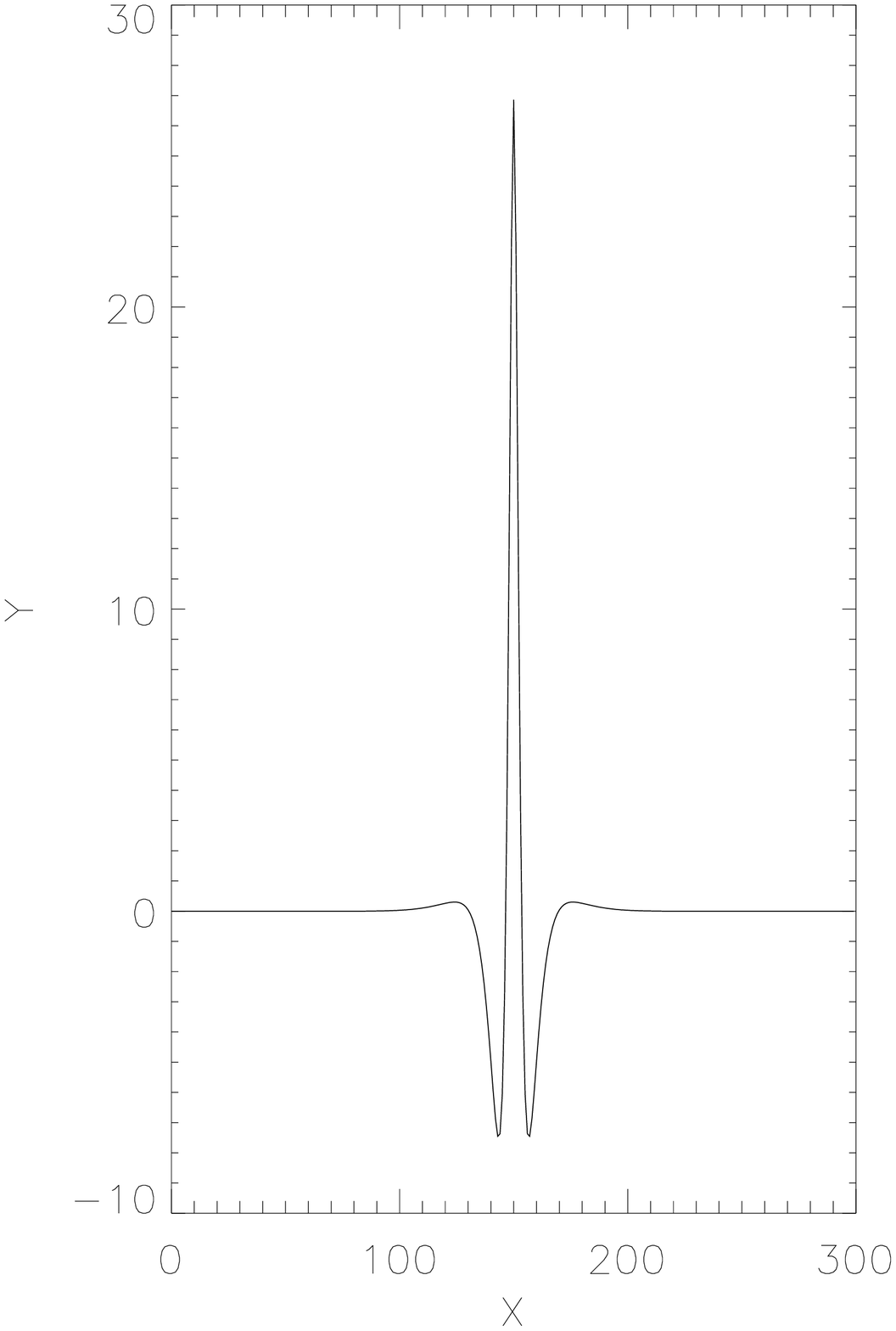}
\includegraphics[width=0.495\textwidth,clip=]{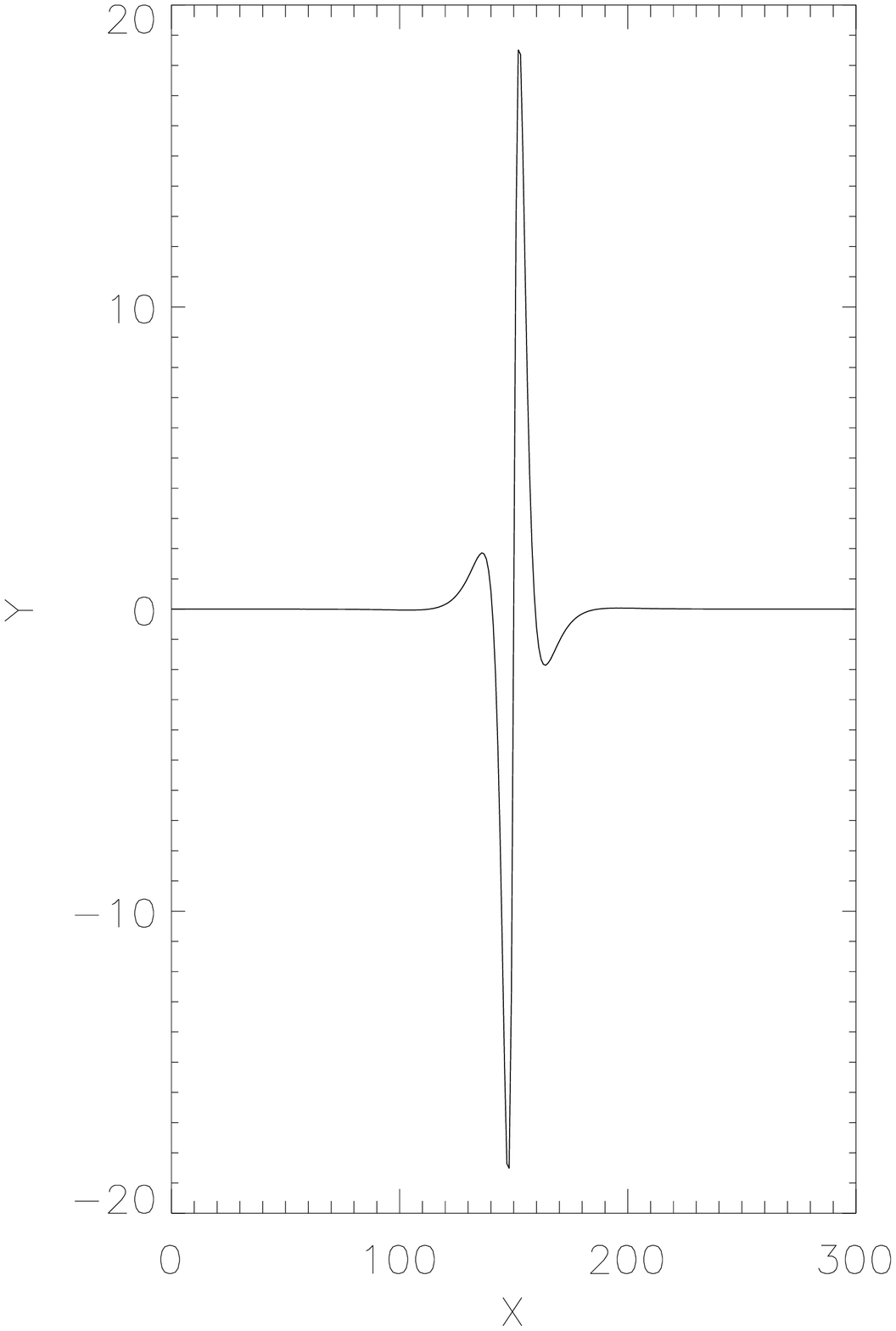}
}
\vspace{-0.58\textwidth}   
\centerline{\Large \bf     
\hspace{0.195\textwidth} \color{black}{(a)}
\hspace{0.385\textwidth} \color{black}{(b)}
\hfill}
\vspace{0.54\textwidth}    
\caption{An example of Log-Gabor functions [$Y = G(X)$] in the spatial domain, which are obtained by a numerical inverse Fourier Transform of~\autoref{eq:LogGabor}. (a): Real part. (b): Imaginary part}
\label{fig:LogGabor}
\end{figure}

Phase congruency is a rather awkward quantity to calculate and the Fourier transform is not good for determining local frequency information. As an alternative, \citet{Kovesi1995} describes a way of calculating phase congruency using Log-Gabor wavelets. The Log-Gabor function is proposed by \citet{Field1987}, who suggests that natural images are better coded by filters that have Gaussian transfer functions when viewed on the logarithmic frequency scale. The transfer function can be constructed as 
\begin{equation}
\label{eq:LogGabor}
F= \exp -\frac{\ln^2(\sqrt{u^2_1+u^2_2}/\omega_0)}{2\ln^2(k/\omega_0)}
\end{equation}
where $\omega_0$ is the filter's centre frequency, $k/\omega_0$ controls the filter's bandwidth. Due to the singularity in the log function at the origin, the corresponding functions in the spatial domain, which are a pair of filters in quadrature (see Figure~\ref{fig:LogGabor}), can be obtained by a numerical inverse Fourier Transform. The appearance is similar to Gabor functions, although their shape becomes much sharper as the bandwidth is increased. Therefore, Log-Gabor wavelets can be constructed with arbitrary bandwidth and the bandwidth can be optimised to produce a filter with minimal spatial extent \citep{KovesiCode}.

\begin{figure}    
\centerline{
\includegraphics[width=0.75\textwidth,clip=]{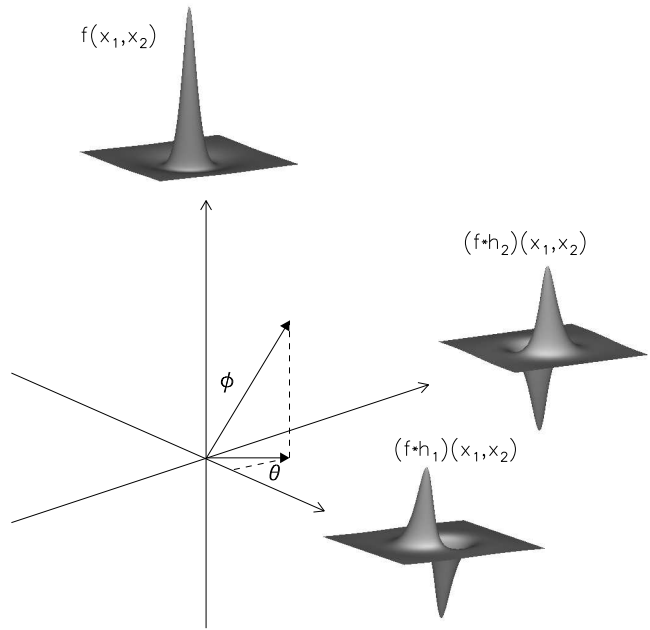}
}
\caption{Local phase and amplitude of monogenic filters, constructed following  \citet{Kovesi2012}.}
\label{fig:PCM}
\end{figure}

\citet{Kovesi1995} extends the one-dimensional Log-Gabor filters into two dimensions by  applying a Gaussian spreading with respect to several preference directions. In order to circumvent the complexity of combining phase congruency information over different orientations, \citet{Felsberg2000} presents a vector-valued filter that is odd and has an isotropic energy distribution; then the whole theory of local phase and amplitude can directly be applied to images. In the frequency domain, this filter can be formed by combining the Log-Gabor filter with its Riesz transform
\begin{eqnarray}
H_1F&=& F\times i\frac{u_1}{\sqrt{u^2_1+u^2_2}}\\
H_2F&=& F\times i\frac{u_2}{\sqrt{u^2_1+u^2_2}}\nonumber
\end{eqnarray}
where $F$ is the Log-Gabor transfer function (\autoref{eq:LogGabor}), $(u_1,u_2)$ are the coordinates in the frequency domain, and $i$ is the imaginary unit. The vector $\vec{H}=(H_1,H_2)$ is isotropic and odd because $\sqrt{H_1^2+H_2^2}=1$ and $\vec{H}(-u_1,-u_2)=-\vec{H}(u_1,u_2)$. These two filters represent a quadrature phase shifting operation in the two orthogonal directions of the image \citep{Kovesi2012}. To obtain local phase and amplitude information, the image $I$ is convolved with the Log-Gabor filter $f$ and the two Reisz transform filtered version of $f$, $h_1f$, and $h_2f$. Then the triple $(I*f,I*h_1f,I*h_2f)$ forms a multi-dimensional generalization of the analytic signal which is called the monogenic signal, where $*$ denotes convolution. The local amplitude of the monogenic signal is
\begin{equation}
A = \sqrt{(I*f)^2+(I*h_1f)^2+(I*h_2f)^2}
\end{equation}
and the local phases can be represented in the standard spherical coordinates
\begin{eqnarray}
     I*f&=&A\cos(\phi)\nonumber\\
(I*h_1f)&=&A\sin(\phi)\cos(\theta)\\
(I*h_2f)&=&A\sin(\phi)\sin(\theta)\nonumber
\end{eqnarray}
Figure~\ref{fig:PCM} depicts this process. The output from convolution with the Log-Gabor filter $f$ corresponds to the real component of the analytic signal and the convolutions with the Reisz tranform filters [$(h_1f,h_2f)$] corresponds to the imaginary component of the analytic signal. Eventually, phase [$\phi$] and amplitude [$A$] at multi-scales are used to calculate the phase congruency through \autoref{eq:PC}.

The monogenic signal also contains information about the local orientation of an image, namely, phase $\theta$. Therefore, it can be used to perform Nonmaxima Suppression as Canny edge-detection algorithm \citep{Canny1986}. The Canny edge-detection algorithm uses a pair of convolution masks to estimate the gradients of an image in the x-direction and y-direction, then computes the gradient orientation and magnitude. An edge point is defined to be a point whose gradient magnitude is locally maximum along the direction of the gradient. This process, which results in ridges one pixel wide, is called nonmaxima suppression. As an analogy, we can constrain the phase congruency to a local maxima across the direction defined by the local orientation [$\theta$]. The Canny edge-detection algorithm also uses a hysteresis thresholding to process the nonmaxima suppression image. Hysteresis thresholding adopts two thresholds [$T_1$ and $T_2$]: any pixel in the nonmaxima suppression image that has a value greater than $T_1$ is presumed to be an edge, and any pixels that are connected to this edge pixel and have a value greater than $T_2$ are also selected as edge pixels.


\end{article} 

\end{document}